\documentclass[aps, prd, superscriptaddress, preprintnumbers, showpacs]{revtex4}

\usepackage{graphicx,amsfonts,amsmath,amssymb,amstext}
\usepackage{float,wrapfig}
\usepackage{subfigure, psfrag}
\usepackage{dsfont}

\usepackage{color}
\usepackage{bm}

\newcommand{\be}{\begin{equation}}
\newcommand{\ee}{\end{equation}}
\newcommand{\ba}{\begin{eqnarray}}
\newcommand{\ea}{\end{eqnarray}}

\newcommand{\tr}{\,\mbox{tr}}
\newcommand{\sign}{\,\mbox{sign}}

\definecolor{purple}{rgb}{0.8,0,0.6}
\definecolor{darkgreen}{rgb}{0.00,0.6,0.00}

\begin{document}

\title{Electrified magnetic catalysis in three-dimensional topological insulators}
\date{\today}

\author{E.~V.~Gorbar}
%\email{gorbar@bitp.kiev.ua}
\affiliation{Department of Physics, Taras Shevchenko National Kiev University, Kiev, 03680, Ukraine}
\affiliation{Bogolyubov Institute for Theoretical Physics, Kiev, 03680, Ukraine}

\author{V.~A.~Miransky}
%\email{vmiransk@uwo.ca}
\affiliation{Department of Applied Mathematics, Western University, London, Ontario N6A 5B7, Canada}
\affiliation{Department of Physics and Astronomy, Western University, London, Ontario N6A 3K7, Canada}

\author{I.~A.~Shovkovy}
%\email{igor.shovkovy@asu.edu}
\affiliation{College of Letters and Sciences, Arizona State University, Mesa, Arizona 85212, USA}

\author{P.~O.~Sukhachov}
%\email{psukhach@uwo.ca}
\affiliation{Department of Applied Mathematics, Western University, London, Ontario N6A 5B7, Canada}
\affiliation{Department of Physics, Taras Shevchenko National Kiev University, Kiev, 03680, Ukraine}

\begin{abstract}
The gap equations for the surface quasiparticle propagators in a slab of three-dimensional topological
insulator in external electric and magnetic fields perpendicular to the slab surfaces are analyzed and
solved. A different type of magnetic catalysis is revealed with the dynamical generation of {\it both} Haldane
and Dirac gaps. Its characteristic feature manifests itself in the crucial role that the electric field plays
in dynamical symmetry breaking and the generation of a Dirac gap in the slab. It is argued
that, for a sufficiently large external electric field, the ground state of the system is a phase with a
homogeneous surface charge density.
\end{abstract}

%71.70.Di Landau levels
%73.20.-r Electron states at surfaces and interfaces
%73.22.Gk Broken symmetry phases
%71.10.Pm   Fermions in reduced dimensions (anyons, composite fermions, Luttinger liquid, etc.)

\maketitle

\section{Introduction}
\label{sec:Introduction}

Topological insulators (TIs) form a class of materials with unique properties, associated with a non-trivial
topology of their quasiparticle band structure (for a review, see
Refs.~\cite{Zhang:rev, Hasan-Kane:rev, Hasan-Moore:rev, Ando:rev}).
The key feature of two-dimensional (2D) and three-dimensional (3D) TIs is the
existence of special gapless edge and surface states, respectively, while the bulk states of those materials
are gapped. The hallmark property of the surface states is their topological protection. Mathematically, the
nontrivial topological properties of time-reversal (TR) invariant TIs are generally described \cite{Moore:2007}
by multiple copies of the $Z_2$ invariants found by Kane and Mele \cite{Kane-Mele}. This implies
that the energy band gap should close at the boundary between topological and trivial insulator
(e.g., vacuum) giving rise to the occurrence of the gapless interface states and the celebrated bulk-boundary
correspondence. The discovery of the $Z_2$ topology in TIs is an important breakthrough
because it showed that nontrivial topology can be embedded in the band structure and that the presence
of an external magnetic field is not mandatory for the realization of topological phases.

Another distinctive feature of the 3D TIs is a relativistic-like energy spectrum of the surface states, whose
physical origin is related to a strong spin-orbit coupling \cite{Hsieh:2009}. Indeed, the surface states on
each of the surfaces are described by 2D massless Dirac fermions in an irreducible  2$\times$2 representation,
with a single Dirac point in the reciprocal space. For comparison, quasiparticles in graphene demonstrate
similar properties, but have four inequivalent Dirac cones due to a spin and valley degeneracy \cite{Castro:2009}
that makes certain aspects of their physics very different from those of the surface states in TIs. In
our study below, we will concentrate only on the case of the strong 3D TIs whose surface states are protected
by the topology of the bulk bands in combination with the TR symmetry. This leads to the locking of
momenta and spin degrees of freedom and, consequently, to the formation of a helical Dirac (semi)metal state
\cite{Hsieh:2009}. Such a state is characterized by the electron antilocalization and the absence
of backscattering. The phenomenon of antilocalization has deep mathematical roots and
is usually explained by an additional Berry's phase $\pi$ that is acquired when an electron circles a Dirac
point. From the physical viewpoint, when scattering on an impurity, an electron must change its spin in
order to preserve its chirality. Such a process is possible only in the case of magnetic impurities which break
explicitly the
TR symmetry.

Experimentally, a linear relativistic-like dispersion law of the surface states is observed in Bi$_{1-x}$Sb$_x$,
Bi$_2$Se$_3$, Bi$_2$Te$_3$, Sb$_2$Te$_3$, Bi$_2$Te$_2$Se, and other materials by using angle resolved
photoemission spectroscopy (ARPES) \cite{Hsieh:2008, Zhang:2009, Hsieh:2009, Chen:2009, Cava-Hasan}.
Furthermore, scanning tunneling microscopy and scanning tunneling spectroscopy provide additional
information about the topological nature of the surface states, such as the quasiparticles interference
patterns around impurities and defects. The Fourier analysis of these patterns has shown that the
backscattering between $\mathbf{k}$ and $-\mathbf{k}$ is
highly suppressed in Bi$_{1-x}$Sb$_x$ \cite{Roushan:2009} and Bi$_2$Te$_3$ \cite{Zhang-Cheng:2009}
in accord with the TR symmetry protection. The existence of an odd number of Dirac nodes
leads to other exotic properties associated with surface states of TIs, e.g., an axion electromagnetic response
\cite{Qi:2008}, an unusual surface Hall conductance \cite{Fu:2007-76,Xu:2014}, etc.

It is well known that electrons confined to two dimensions can form numerous interaction-induced phases.
By using numerical calculations, it was shown in Ref.~\cite{Koulakov:1996} that it is energetically favorable
for the 2D electron liquid in a weak magnetic field to form domains with empty and fully filled higher Landau
levels. Depending on the number of Landau levels filled, the corresponding charge density wave
(CDW) phase is realized with a ``stripe'' or ``bubble'' pattern. By using the simplest model of the
surface states in a magnetic field with strong local repulsion and a long-range Coulomb interaction
included perturbatively, it was suggested that a similar CDW phase with a ``stripe'' or ``bubble'' pattern
can be also realized on the surface of 3D TIs for supercritical values of a local repulsion constant
\cite{Vishwanath:2010}. For subcritical local repulsion, the composite Fermi liquid (CFL) \cite{HLR}
phase is expected \cite{Rezayi:2000, Vishwanath:2010}. It is interesting that composite fermions
in conventional 2D electron gas at half-filling were recently suggested to be massless Dirac (quasi)particles
\cite{Son:2015} similar to the surface quasiparticles of TIs. This result was also checked
numerically in Ref.~\cite{Geraedts:2016}, where it was shown that at the half-filling
the particle-hole symmetry for composite fermions plays the same role as the TR symmetry for
the 2D Dirac fermions and, consequently, the backscattering off symmetry preserving impurities
is also forbidden.

The influence of an external electric field on the exciton condensation in thin films of TIs was
studied in Refs.~\cite{Franz:2009, Efimkin:2012}, where it was shown that the electron condensate effectively
joins the surfaces of a thin film and leads to the formation of a pairing gap. However, this is
important only in thin ($l_z \lesssim 8~\mbox{nm}$) films of TIs and can be ignored in
sufficiently thick slabs \cite{Linder:2009}. The exciton condensate exhibits unusual properties
including a stable zero mode and a fractional charge $\pm e/2$ carried by a singly quantized
vortex in the exciton condensate \cite{Franz:2009}. The dynamical gap
generation in a simple model of TIs was also considered in Ref.~\cite{Cea:2016}.

Just like a magnetic field, an external electric field may play an important role in the dynamics of the
surface states in a 3D TI slab. In this paper, we study the dynamical gap generation and the phase
diagram of a TI slab placed in external magnetic and electric fields perpendicular to the slab surfaces.
(Note that the case of the parallel fields is rather trivial. While a parallel magnetic field does
not affect the orbital motion, a parallel electric field produces a current on the surface.) We
argue that a uniform phase with both dynamically generated Dirac and Haldane gaps is realized
in sufficiently strong (weak) electric (magnetic) fields. Although the explicit calculations performed
in this paper use the model parameters suitable for Bi$_2$Se$_3$, the main qualitative conclusions
should be valid for all similar TIs.

The paper is organized as follows. The effective Hamiltonian of the surface states in the simplest
model of a topological insulator with short- and long-range interactions is described in
Sec.~\ref{sec:model-all}. The set of gap equations at finite temperature is derived in
Sec.~\ref{sec:gap-equation-Coulomb} and its solutions in electric and magnetic fields are
obtained numerically in Sec.~\ref{sec:gap-equation-Coulomb-results-kappa}.
The qualitative description of the inhomogeneous phase with two stripes is given in Sec.~\ref{sec:stripe}.
The main results are discussed and summarized in Secs.~\ref{sec:Discussion} and \ref{sec:Conclusion}, respectively.

For convenience, throughout the paper, we set $\hbar=c=1$.

\section{Model}
\label{sec:model-all}

By projecting the 3D bulk Hamiltonian onto the subspace of surface states (see Refs.~\cite{Zhang:rev, Shen:book} for a detailed
consideration), the following effective Hamiltonian for the top surface of a 3D TI is obtained
\cite{Zhang:2009,Shan:2010, Zhang:rev,Shen:book}:
\begin{eqnarray}
\mathcal{H}_{\mathrm{top}\,\,\mathrm{surf}}(\mathbf{k})=C+v_F\left(\bm{\sigma}\times\mathbf{k}\right)_z +\mathcal{O}\left(\mathbf{k}^2\right)
= C+\left(
                               \begin{array}{cc}
                                 0 & v_F(ik_x+k_y) \\
                                 v_F(-ik_x+k_y) & 0 \\
                               \end{array}
                             \right)+ \mathcal{O}\left(\mathbf{k}^2\right),
\label{model-H-2D}
\end{eqnarray}
where $C$ is a constant, $\bm{\sigma}=\left(\sigma_x, \sigma_y\right)$ are the Pauli matrices,
$v_F=4.1~\mbox{eV\AA}=6.2\times10^5~\mbox{m/s}$ is the Fermi velocity, and
$\mathbf{k}=\left(k_x, k_y\right)$ is the surface momentum. The effective surface Hamiltonian
for the bottom surface is obtained from the Hamiltonian of the top surface by the inversion $\mathbf{k} \to
-\mathbf{k}$ (see Sec.~III.C in Ref.~\cite{Zhang:rev}). It is worth noting that the effective surface Hamiltonian is valid only at
sufficiently small chemical potentials, when the bulk states are gapped. Therefore, the corresponding energy cutoff
can be approximated by the bulk band gap, i.e., $\Lambda\simeq \Delta_{\rm bulk}$. In the case of Bi$_2$Se$_3$,
for example, $\Delta_{\rm bulk} \approx 0.35~\mbox{eV}$ \cite{Mooser, Black}.

The resulting model Hamiltonian, describing quasiparticle states on the top and bottom surfaces of the
3D TI in constant electric and magnetic fields applied perpendicular to the surfaces of the slab, is
given by $H^{(0)}=H^{(0)}_{+}\oplus H^{(0)}_{-}$, where
\begin{equation}
H^{(0)}_{\lambda}=\int d^2\mathbf{r}\,\,\psi^{\dag}_{\lambda}(\mathbf{r})\left(
                        \begin{array}{cc}
                         m^{(0)} -\mu_{\lambda}^{(0)} & iv_F(\pi_x-i\pi_y) \\
                          -iv_F(\pi_x +i\pi_y) & -m^{(0)} -\mu_{\lambda}^{(0)} \\
                        \end{array}
                      \right)\psi_{\lambda}(\mathbf{r}).
\label{model-H-s-0}
\end{equation}
Here $\lambda=\pm$ denotes the top and bottom surfaces, respectively, $\mu_{\lambda }^{(0)}$
is the surface electrochemical potential, $\bm{\pi} \equiv -i \bm{\nabla} + e\mathbf{A}$ is the canonical
momentum, $\mathbf{A}=\left(0, Bx\right)$ is the vector potential that describes the constant magnetic field
$\mathbf{B}$ pointing in the $z$ direction, and $e$ is the electron charge. Note that
in Eq.~(\ref{model-H-s-0}) we redefined the wave function on the bottom surface by
replacing $\psi_{-}\to\sigma_z\psi_{-}$. As the notation suggests, the value of $\mu^{(0)}_{\lambda}$ may
depend on the surface index $\lambda$. Indeed, this is quite natural
in the model at hand since fixing charge densities on the top and bottom surfaces requires an introduction of
the corresponding local electrochemical potentials. In view of a large surface $g$-factor, $g_{s}=18\pm4$
\cite{Fu:2016}, the Zeeman splitting is important in TIs. This spin splitting is included in Hamiltonian
(\ref{model-H-s-0}) as the bare gap parameter $m^{(0)} =g_s\mu_{B} B/2$, where $\mu_{B}=5.788
\times 10^{-5}~\mbox{eV/ T}$ is the Bohr magneton.

Before proceeding with the analysis of the model, it is convenient to rewrite the model Hamiltonian (\ref{model-H-s-0})
in terms of the Dirac matrices. It is well known that there are two irreducible representations of the Clifford-Dirac
algebra in (2+1)-dimensions, e.g., see Ref.~\cite{Appelquist:1986}. One of them is
\begin{equation}
\gamma^0=\sigma_z, \quad \gamma^1=i\sigma_x, \quad \gamma^2=i\sigma_y
\label{model-gamma-matrices}
\end{equation}
and the other irreducible representation is obtained by changing $\gamma^{\mu}\to-\gamma^{\mu}$
with $\mu=0, 1, 2$ in Eq.~(\ref{model-gamma-matrices}). In terms of the Dirac matrices
(\ref{model-gamma-matrices}), the free Hamiltonian (\ref{model-H-s-0}) takes the following form:
\begin{equation}
H^{(0)}_{\lambda}=\int d^2\mathbf{r}\,\,\bar{\psi}_{\lambda}(\mathbf{r})\left(
                       -\mu^{(0)}_{\lambda}\gamma^0+v_F(\bm{\pi}\cdot\bm{\gamma})+m^{(0)}
                      \right)\psi_{\lambda}(\mathbf{r}),
\label{model-H-0-matrices}
\end{equation}
where $\bar{\psi}_{\lambda}(\mathbf{r})=\psi^{\dag}_{\lambda}(\mathbf{r})\gamma^0$.
When an external electric field is applied perpendicularly to the surfaces of the TI slab, the gapless
surface states will tend to completely screen the field out. Indeed, from a physics viewpoint, the TI slab is like
a Faraday cage made of gapless (metallic) surface states enclosing a gapped (insulating) interior. This implies
that there should be no electric field inside a (sufficiently thick) TI slab. Enforcing this condition allows one
to determine the charge densities and electrochemical potentials on the surfaces. In terms of the charge densities
on the top and bottom surfaces, one has
\begin{equation}
\rho_{\lambda }=\lambda \epsilon_0 \mathcal{E},
\label{model-DOS-zero-field}
\end{equation}
where $\mathcal{E}$ is the external electric field pointing in the $z$-direction, $\epsilon_0 \approx 8.854
\times10^{-12}~\mbox{F/m}$ is the permittivity of free space, and $2\epsilon_0 \mathcal{E}$
corresponds to the difference of the charge densities of the top and bottom surfaces needed to compensate
the external electric field.

Under the parity transformation $P$ in (2+1) dimensions, which changes
the sign of a spatial coordinate, i.e., $(x,y) \to (-x,y)$, the two-component spinors transform
as follows: $P\psi(t,x,y)P^{-1}=\sigma_x\psi(t,-x,y)$. Clearly, the last term in Hamiltonian (\ref{model-H-0-matrices})
breaks parity, as well as the TR symmetry $T\psi(t,x,y)T^{-1}=\sigma_y\psi^{*}(-t,x,y)$. This mass term
is known in the literature as the Haldane mass $\sum_{\lambda}m_H\,\bar{\psi}_{\lambda}\psi_{\lambda}$ \cite{Haldane}.
A parity and TR invariant mass is also possible in the model with two irreducible representations.
It is given by the Dirac mass term $\sum_{\lambda}m_D\,\lambda\bar{\psi}_{\lambda}\psi_{\lambda}$
with the parity transformation defined by $\psi_{\lambda=+1} \to \sigma_x \psi_{\lambda=-1}$ and
$\psi_{\lambda=-1} \to \sigma_x \psi_{\lambda=+1}$. (Note that, in the TI slab model, this
transformation interchanges the states on the different spatially separated surfaces.)
While a Chern--Simons mass term for the gauge field is induced via one-loop polarization when the Haldane
mass is present, the Chern-Simons term is absent in the case of the Dirac mass. The spontaneous breaking
of parity in $(2+1)$-dimensional QED was studied in Ref.~\cite{Appelquist1:1986}.

In this study, the model interaction Hamiltonian $H_{\rm int}$ includes both a long-range Coulomb
and a short-range local four-fermion interactions
\begin{equation}
H_{\rm int} = \frac{e^2}{8\pi \epsilon_0 \kappa_{\rm surf}}\int d^2\mathbf{r}d^2\mathbf{r}^{\prime}\,\frac{\Psi^{\dagger}(\mathbf{r})
\Psi(\mathbf{r})
\Psi^{\dagger}(\mathbf{r}^{\prime})\Psi(\mathbf{r}^{\prime})}{|\mathbf{r}-\mathbf{r}^{\prime}|} +\frac{G_{\rm int}}{2}\int d^2\mathbf{r}\,
\Psi^{\dagger}(\mathbf{r})\Psi(\mathbf{r})
\Psi^{\dagger}(\mathbf{r})\Psi(\mathbf{r}),
\label{model-H-int}
\end{equation}
where $\Psi(\mathbf{r})=\left(\psi_{\lambda=+1}(\mathbf{r}),\psi_{\lambda=-1}(\mathbf{r})\right)^T$. The first term
in $H_{\rm int}$ describes the long-range Coulomb interaction and takes into account the effective surface
dielectric constant $\kappa_{\rm surf}=\left(1+\kappa_{\rm bulk}\right)/2 \approx 56$, where the bulk dielectric
constant $\kappa_{\rm bulk}\approx113$ for Bi$_2$Se$_3$ \cite{Madelung}. The second term captures the
on-site local repulsion, parametrized by the dimensionful coupling constant $G_{\rm int}$. In view of the large bulk dielectric
constant and assuming a large slab thickness, we neglect the intersurface interaction and the possible formation of
an intersurface exciton condensate \cite{Franz:2009, Efimkin:2012}. Thus, the full Hamiltonian of our
model is given by the sum of the free and interaction Hamiltonians in Eqs.~(\ref{model-H-0-matrices}) and (\ref{model-H-int}).

\section{Gap equations}
\label{sec:gap-equation-Coulomb}

In this section, we study the gap generation in the effective model of a sufficiently thick TI slab
described in the previous section. The inverse free surface fermion propagator is given by
\begin{equation}
iS^{-1}_{\lambda}(u,u^\prime) = \left[(i\partial_t+\mu^{(0)}_{\lambda })\gamma^0
-v_F(\bm{\pi}\cdot\bm{\gamma})-m^{(0)}\right]\delta^{3}(u-u^\prime),
\label{gapEq-Coulomb-sinverse}
\end{equation}
where $u=\left(t, \mathbf{r}\right)$ denotes a space-time coordinate. By using this as a guide,
we assume the following rather general ansatz for the inverse full surface fermion propagator:
\begin{eqnarray}
iG^{-1}_{\lambda}(u,u^\prime)= \left[(i\partial_t+\mu_{\lambda })\gamma^0
-v_F(\bm{\pi}\cdot\bm{\gamma})-m_{\lambda}\right]\delta^{3}(u-u^\prime),
\label{gapEq-Coulomb-ginverse}
\end{eqnarray}
where $m_{\lambda }$ is a dynamically generated gap (mass) which, in general, includes both
Haldane and Dirac gaps and $\mu_{\lambda }$ denotes the dynamical electrochemical potential.
Note that all dynamical parameters in the full propagator are assumed to be functions of
$(\bm{\pi}\cdot\bm{\gamma})^2l^2$, where $l = 1/\sqrt{|eB|}$ is the magnetic length. Therefore, in
the end, they all depend on the Landau level index $n$.

Because of the long-range interaction, in principle, the renormalization of the wave function
should be included in the full propagator (\ref{gapEq-Coulomb-ginverse}). This can be formally done
by replacing the Fermi velocity $v_F$ with a dynamical function $F_{\lambda}$. It is well justified,
however, to neglect the renormalization of the Fermi velocity and replace it with $v_F$. Indeed, even
in the case of graphene with an unscreened Coulomb interaction, the renormalized Fermi velocity
is generically $10\%$ to $30\%$ larger than the corresponding bare value $v_F$
\cite{2011NatPh7,Gonzalez:1993uz,Gorbar:2011kc}. Because
of a much larger surface dielectric constant and, consequently, a much smaller coupling constant,
the Coulomb interaction will play a minor role in the Fermi velocity renormalization and,
as we will show below, in the generation of dynamical gaps in TIs.

In order to represent the inverse propagator in the form of a Landau-level expansion, we use the
following complete set of eigenstates (for details, see Appendix~A in Ref.~\cite{Gorbar:2011kc}):
\begin{equation}
\psi_{n, k_y}(\mathbf{r})=\frac{1}{\sqrt{2^{n+1} \pi l n!}} H_n\left(k_yl+\frac{x}{l}\right)e^{-\frac{1}{2l^2}\left(x+k_yl^2\right)^2}
e^{is_Bk_yy},
\label{gapEq-Coulomb-psi-nky}
\end{equation}
where $H_n(x)$ are the Hermite polynomials and $s_{B}=\mbox{sign}(eB)$. By making use of the results in Appendix~\ref{sec:wf-Green},
we derive the following inverse fermion propagators
in the mixed frequency-momentum representation:
\begin{eqnarray}
\label{gapEq-Coulomb-sinverse-LL-no-phase}
S^{-1}_{\lambda}(\omega,\mathbf{r},\mathbf{r}^\prime)&=& e^{i\Phi(\mathbf{r},\mathbf{r}^\prime)}\tilde{S}_{\lambda}^{-1}(\omega,\mathbf{r}-\mathbf{r}^\prime) ,\\
\label{gapEq-Coulomb-ginverse-LL-no-phase}
G^{-1}_{\lambda}(\omega,\mathbf{r},\mathbf{r}^\prime)&=& e^{i\Phi(\mathbf{r},\mathbf{r}^\prime)}\tilde{G}_{\lambda}^{-1}(\omega,\mathbf{r}-\mathbf{r}^\prime).
\end{eqnarray}
Here $\Phi(\mathbf{r}, \mathbf{r}^{\prime})= -eB(x+x^{\prime})(y-y^{\prime})/2$ is the famous Schwinger phase
and the translation invariant parts of the inverse propagators are given by
\begin{eqnarray}
\label{gapEq-Coulomb-sinverse-LL}
iS^{-1}_{\lambda}(\omega,\mathbf{r}-\mathbf{r}^\prime)&=& \frac{e^{-\eta/2}}{2\pi l^2}\sum_{n=0}^{\infty}
\Big\{ s_B (\omega+\mu^{(0)}_{\lambda })\left[P_{+} L_{n-1}(\eta)-P_{-} L_{n}(\eta)\right]-m^{(0)}\left[P_{+} L_{n-1}(\eta)+P_{-} L_{n}(\eta)
\right] \nonumber \\
&-&\frac{i}{l^2}v_F (\bm{\gamma}\cdot\mathbf{r}) L^1_{n-1}(\eta)\Big\},\\
\label{gapEq-Coulomb-ginverse-LL}
iG^{-1}_{\lambda}(\omega,\mathbf{r}-\mathbf{r}^\prime)&=& \frac{e^{-\eta/2}}{2\pi l^2}\sum_{n=0}^{\infty}
\Big\{ s_B (\omega+\mu_{n,\lambda })\left[P_{+} L_{n-1}(\eta)-P_{-} L_{n}(\eta)\right]-m_{n,\lambda }\left[P_{+} L_{n-1}(\eta)+P_{-} L_{n}(\eta)
\right]\nonumber \\
&-&\frac{i}{l^2}v_F (\bm{\gamma}\cdot\mathbf{r}) L^1_{n-1}(\eta)\Big\},
\end{eqnarray}
where $P_{\pm} =\left(1 \pm s_B \gamma^0\right)/2$,
$\eta= (\mathbf{r}-\mathbf{r}^{\prime})^2/(2l^2)$, and
$L^{j}_{n} \left(x\right)$ are the generalized Laguerre polynomials (by definition $L_n\equiv L_n^0$).

In order to study the dynamical gap generation, we utilize the Baym--Kadanoff (BK) formalism \cite{BK},
which leads to a self-consistent Schwinger-Dyson equation for the fermion propagator. In contrast to a
perturbative analysis, the BK formalism can capture nonperturbative effects such as spontaneous symmetry breaking.
To leading order in coupling, the BK effective action in the model under consideration is given by Eq.~(\ref{app-action-BK})
in Appendix~\ref{sec:App-action-true}. In view of the geometry of conducting states of our TI system, it should
not be too surprising that the effective action (\ref{app-action-BK}) has a form similar to that in bilayer graphene
(compare with Eq.~(9) in Ref.~\cite{bilayer}).

The extremum of the effective action $\frac{\delta \Gamma(G)}{\delta G_{\lambda}} =0$ defines the following
Schwinger-Dyson equation for the full fermion propagator (for details, see Appendix~\ref{sec:App-action-true}):
\begin{equation}
iG^{-1}_{\lambda}(u,u^\prime) = iS^{-1}_{\lambda}(u,u^\prime)
- e^2  \gamma^0 G_{\lambda}(u,u^{\prime}) \gamma^0 D(u^{\prime}-u)
-G_{\rm int}\left\{ \gamma^0 G_{\lambda}(u,u) \gamma^0 - \gamma^0\, \mbox{tr}[\gamma^0G_{\lambda}(u,u)]\right\}\delta^{3}(u-u^{\prime}),
\label{gapEq-Coulomb-gap}
\end{equation}
where the trace in the last term is taken over the spinor indices and the Hartree term due to the Coulomb interaction
is absent. This is justified because of the overall neutrality of the sample, i.e.,
\begin{equation}
Q_b- e\, \sum_{\lambda=\pm}\mbox{tr}[\gamma^0G_{\lambda}(u,u)]=0,
\label{gapEq-Coulomb-Hartree}
\end{equation}
where $Q_b$ denotes the background charge due to the external gates. We note, however, that it does
not make sense to drop the Hartree-type term due to the contact interaction. Therefore, the
corresponding term is kept in curly brackets in Eq.~(\ref{gapEq-Coulomb-gap}).

The propagator mediating the Coulomb interaction is denoted by $D(u)$. Its explicit expression is given by
\begin{equation}
D(u)=\int\frac{d\omega d^2\mathbf{k}}{(2\pi)^3}D(\omega, \mathbf{k})e^{-i\omega t+i\mathbf{k}\cdot\mathbf{r}}\approx\delta(t)
\frac{1}{4\pi\epsilon_0\kappa_{\rm surf}} \int \frac{dk}{2\pi} \frac{kJ_0(kr)}{k+\Pi(0,k)},
\label{gapEq-Coulomb-D}
\end{equation}
where $J_0(x)$ is the Bessel function. In the last expression, we neglected the dependence of the
polarization function $\Pi(\omega,k)$ on $\omega$, which corresponds to an instantaneous approximation.
Such an approximation may be reasonable for the TI surfaces, where charge carriers propagate
much slower than the speed of light and, thus, the retardation effects are negligible. It is worth noting,
however, that the instantaneous approximation has a tendency to underestimate the strength
of the Coulomb interaction \cite{Gorbar:2002iw}.

Just like the inverse propagators in Eqs.~(\ref{gapEq-Coulomb-sinverse-LL-no-phase}) and
(\ref{gapEq-Coulomb-ginverse-LL-no-phase}), the propagators themselves have the same
Schwinger phase. The full propagator, in particular, takes the following explicit form:
\begin{eqnarray}
\label{gapEq-Coulomb-G-phase}
G_{\lambda}(\omega, \mathbf{r}, \mathbf{r}^{\prime}) &=& e^{i\Phi(\mathbf{r},\mathbf{r}^\prime)}\tilde{G}_{\lambda}(\omega, \mathbf{r}-\mathbf{r}^{\prime}) ,\\
\label{gapEq-Coulomb-G-no-phase}
\tilde{G}_{\lambda}(\omega, \mathbf{r}-\mathbf{r}^{\prime})&=& \frac{e^{-\eta/2}}{2\pi l^2}\sum_{n=0}^{\infty}
\Bigg\{\frac{s_B\left(\omega
+\mu_{n,\lambda } \right)\left[ L_{n-1}(\eta)P_{+} -L_{n}(\eta)P_{-}\right]}{\left(\omega+\mu_{n,\lambda }
+i0\sign{(\omega)}\right)^2-M_n^2} \nonumber\\
&+&\frac{m_{n,\lambda } \left[L_{n-1}(\eta)P_{+} +L_{n}(\eta)P_{-}\right] -i\frac{v_F}{l^2}L_{n-1}^1(\eta)
\left(\bm{\gamma}\cdot(\mathbf{r}-\mathbf{r}^{\prime})\right) }{\left(\omega+\mu_{n,\lambda } +i0\sign{(\omega)}\right)^2-M_n^2}
\Bigg\},
\end{eqnarray}
where $M_n=\sqrt{\left(m_{n, \lambda }\right)^2+\epsilon_{B}^2n}$ and $\epsilon_{B}= \sqrt{2v_F^2|eB|}$ is the Landau energy scale.
The inverse and full fermion propagators at finite temperature are easily obtained through the standard replacement
$\omega\to i\omega_{m^{\prime}}=i\pi T(2m^{\prime}+1)$.

By factorizing the Schwinger phase on both sides of Eq.~(\ref{gapEq-Coulomb-gap}), we arrive at
the following gap equation for the translation invariant part of the full propagator:
\begin{equation}
i\tilde{G}^{-1}_{\lambda}(\omega, \mathbf{r}) = i\tilde{S}^{-1}_{\lambda}(\omega, \mathbf{r})
- \alpha v_F \!\! \int\frac{d\Omega}{2\pi} \frac{dk}{2\pi}
\frac{kJ_0(kr)}{k+\Pi(0,k)} \gamma^0 \tilde{G}_{\lambda}(\Omega, \mathbf{r}) \gamma^0
-G_{\rm int}\!\int\frac{d\Omega}{2\pi}  \delta^2(\mathbf{r})
\left\{ \gamma^0 \tilde{G}_{\lambda}(\Omega, \mathbf{r}) \gamma^0 - \gamma^0\,
\mbox{tr}[\gamma^0\tilde{G}_{\lambda}(\Omega, \mathbf{r})]\right\}.
\label{gapEq-Coulomb-gap-1}
\end{equation}
Here we introduced the following notation $\alpha=e^2/(4\pi \epsilon_0 v_F \kappa_{\rm surf})$. In the case of
Bi$_2$Se$_3$, in particular, $\alpha\approx0.062$. Although it is hard to estimate $G_{\rm int}$ reliably, its origin
is the Coulomb repulsion on distance scales comparable to the lattice spacing. Thus, it may be reasonable to use
the following approximate model value:
\begin{equation}
G_{\rm int}=\frac{\alpha v_F^2\kappa_{\rm surf}}{\Delta_{\rm bulk}} \approx 168.7~\mbox{eV\AA}^2,
\label{gapEq-Coulomb-Gint}
\end{equation}
where the factor $\kappa_{\rm surf}$ was introduced in order to compensate for polarization
effects in the definition of $\alpha$. Indeed, polarization effects should be negligible at small distances. It is
worth noting that the corresponding dimensionless constant
\begin{equation}
g_{\rm int}= \frac{G_{\rm int}\Delta_{\rm bulk}}{8\sqrt{2\pi} v_F^2}\approx 0.18
\label{gapEq-Coulomb-g-def}
\end{equation}
is rather small. In fact, it is an order of magnitude smaller than the critical value $g_{cr}=\sqrt{\pi}$ needed for
generating a gap in a (2+1)-dimensional model in the absence of a magnetic field \cite{Gusynin:1994re}. Because
of this and because of the strong suppression of the Coulomb interaction by the large dielectric constant, no dynamical
generation of a gap is expected in such a TI material in the absence of an external magnetic field. Consequently,
the magnetic catalysis \cite{Gusynin:1994re} will play a crucial role in the generation of dynamical gaps in TIs. (For a recent review on magnetic catalysis, see Ref.~\cite{Miransky:2015ava}.)

By using the explicit form of the fermion propagator (\ref{gapEq-Coulomb-G-no-phase}) on the right-hand
side of Eq.~(\ref{gapEq-Coulomb-gap-1}), we can easily calculate the integral over $\Omega$ (or the sum
over the Matsubara frequency at nonzero temperature, see Appendix \ref{sec:Greens-function}).
Afterwards, by multiplying both sides of the gap
equation (\ref{gapEq-Coulomb-gap-1}) by $e^{-\eta/2}L_{n^{\prime}}(\eta)$ or
$e^{-\eta/2}(\bm{\gamma}\cdot\mathbf{r})L_{n^{\prime}}^1(\eta)$ and then integrating over $\mathbf{r}$,
the complete set of equations for the dynamical parameters can be straightforwardly obtained. In
particular, the gap equations for the lowest Landau level (LLL) parameters are given by
\begin{eqnarray}
\Delta_{\rm eff, \lambda }&=&\mu_{\lambda}^{(0)}+s_Bm^{(0)}+ \alpha \frac{v_F}{2l} \Bigg\{
\mathcal{K}^{(0)}_{0,0}\left[ 1-2n_F\left(\Delta_{\mathrm{eff}, \lambda}\right) \right]
-\sum_{n^{\prime}=1}^{\infty}\mathcal{K}^{(0)}_{n^{\prime},0}\Big[ n_F(M_{n^{\prime}}+\mu_{n^{\prime}, \lambda })-n_F(M_{n^{\prime}}
-\mu_{n^{\prime}, \lambda })\nonumber\\
&-& s_B m_{n^{\prime}, \lambda } \frac{1-n_F(M_{n^{\prime}}+\mu_{n^{\prime}, \lambda })
-n_F(M_{n^{\prime}}
-\mu_{n^{\prime}, \lambda })}{M_{n^{\prime}}} \Big]
 \Bigg\} +\frac{G_{\rm int}}{4\pi l^2}\Bigg\{ \sum_{n^{\prime}=1}^{\infty}
\left[n_F\left(M_{n^{\prime}}+\mu_{n^{\prime}, \lambda }\right)-n_F\left(M_{n^{\prime}}-\mu_{n^{\prime}, \lambda }\right)\right]\nonumber\\
&+& \sum_{n^{\prime}=1}^{\infty}
s_B m_{n^{\prime}, \lambda } \frac{1-n_F\left(M_{n^{\prime}}+\mu_{n^{\prime}, \lambda }\right)-n_F
\left(M_{n^{\prime}}-\mu_{n^{\prime}, \lambda }
\right)}{M_{n^{\prime}}} \Bigg\},
\label{gapEq-Coulomb-gap-Delta}
\end{eqnarray}
where $n_{F}(x)=1/\left(e^{x/T}+1\right)$ is the Fermi-Dirac distribution.
Notice that we introduced an effective LLL electrochemical potential $\Delta_{\mathrm{eff}, \lambda}
=\mu_{0, \lambda}+ s_{B}m_{0, \lambda}$ because the LLL parameters $\mu_{0,\lambda}$ and $m_{0,\lambda}$
cannot be unambiguously defined separately and only their combination $\Delta_{\mathrm{eff}, \lambda}$
has a well-defined physical meaning \cite{Gorbar:2008hu,Goerbig}. Similarly, the equations for the
dynamical parameters associated with higher Landau levels read as
\begin{eqnarray}
m_{n, \lambda }&=&m^{(0)}+ s_B \alpha \frac{v_F}{4l} \Bigg\{
\mathcal{K}^{(0)}_{0,n}\left[ 1-2n_F\left(\Delta_{\mathrm{eff}, \lambda}\right) \right]+
 \sum_{n^{\prime}=1}^{\infty}\mathcal{K}^{(0)}_{n^{\prime}-1,n-1}
\Big[ n_F(M_{n^{\prime}}+\mu_{n^{\prime}, \lambda })-n_F(M_{n^{\prime}}-\mu_{n^{\prime}, \lambda }) \nonumber\\
&+&s_B m_{n^{\prime},\lambda } \frac{1-n_F(M_{n^{\prime}}+\mu_{n^{\prime},\lambda })-n_F(M_{n^{\prime}}-\mu_{n^{\prime},\lambda })}
{M_{n^{\prime}}} \Big] \nonumber\\
&-& \sum_{n^{\prime}=1}^{\infty}\mathcal{K}^{(0)}_{n^{\prime},n}\left[ n_F(M_{n^{\prime}}+\mu_{n^{\prime}, \lambda })-n_F(M_{n^{\prime}}
-\mu_{n^{\prime}, \lambda }) -s_B m_{n^{\prime}, \lambda } \frac{1-n_F(M_{n^{\prime}}+\mu_{n^{\prime}, \lambda })-n_F(M_{n^{\prime}}
-\mu_{n^{\prime}, \lambda })}{M_{n^{\prime}}} \right]
 \Bigg\} \nonumber\\
&+& \frac{G_{\rm int}}{8\pi l^2}\left\{ s_B\left[1-2n_F\left(\Delta_{\mathrm{eff}, \lambda}\right)\right] +2\sum_{n^{\prime}=1}^{\infty}
m_{n^{\prime}, \lambda } \frac{1-n_F\left(M_{n^{\prime}}+\mu_{n^{\prime}, \lambda }\right)-n_F\left(M_{n^{\prime}}-\mu_{n^{\prime}, \lambda }
\right)}{M_{n^{\prime}}}\right\},
\label{gapEq-Coulomb-gap-m}
\end{eqnarray}
\begin{eqnarray}
\mu_{n, \lambda }&=&\mu^{(0)}_{\lambda }+\alpha \frac{v_F}{4l} \Bigg\{ \mathcal{K}^{(0)}_{0,n}
\left[ 1-2n_F\left(\Delta_{\mathrm{eff}, \lambda}\right) \right] -\sum_{n^{\prime}=1}^{\infty}\mathcal{K}^{(0)}_{n^{\prime}-1,n-1}
\Big[ n_F(M_{n^{\prime}}+\mu_{n^{\prime}, \lambda })-n_F(M_{n^{\prime}}-\mu_{n^{\prime}, \lambda }) \nonumber\\
&+&s_B m_{n^{\prime}, \lambda } \frac{1-n_F(M_{n^{\prime}}+\mu_{n^{\prime}, \lambda })-n_F(M_{n^{\prime}}-\mu_{n^{\prime}, \lambda })}
{M_{n^{\prime}}} \Big]
 -\sum_{n^{\prime}=1}^{\infty}\mathcal{K}^{(0)}_{n^{\prime},n}\Big[ n_F(M_{n^{\prime}}+\mu_{n^{\prime}, \lambda })-n_F(M_{n^{\prime}}
-\mu_{n^{\prime}, \lambda }) \nonumber\\
&-&s_B m_{n^{\prime}, \lambda } \frac{1-n_F(M_{n^{\prime}}+\mu_{n^{\prime}, \lambda })-n_F(M_{n^{\prime}}-\mu_{n^{\prime}, \lambda })}
{M_{n^{\prime}}} \Big]
 \Bigg\} \nonumber\\
&-&\frac{G_{\rm int}}{8\pi l^2}\left\{ \left[1-2n_F\left(\Delta_{\mathrm{eff}, \lambda}\right)\right] -2\sum_{n^{\prime}=1}^{\infty}
\left[n_F\left(M_{n^{\prime}}+\mu_{n^{\prime}, \lambda }\right)-n_F\left(M_{n^{\prime}}-\mu_{n^{\prime}, \lambda }\right)\right] \right\}.
\label{gapEq-Coulomb-gap-mu}
\end{eqnarray}
The kernel coefficients $\mathcal{K}^{(0)}_{m,n}$ that capture the long-range interaction
effects in the gap equations are defined in Eq.~(\ref{app-K-def}). In this study, for simplicity, we neglect
all screening effects, i.e., we set $\Pi(0,k)=0$. Then, the numerical analysis greatly simplifies because
the coefficients $\mathcal{K}^{(0)}_{m,n}$ can be calculated analytically, see Eq.~(\ref{app-K-Pi0}).

In addition to the gap equation (\ref{gapEq-Coulomb-gap-1}), the constraints for the surface charge densities in
Eq.~(\ref{model-DOS-zero-field}) should be also satisfied. In terms of the model parameters,
the explicit form of the constraint reads as
\begin{equation}
-\frac{e}{4\pi l^2}\left\{ \left[1-2n_F\left(\Delta_{\mathrm{eff}, \lambda}\right)\right]
-2\sum_{n=1}^{\infty} \left[n_F\left(M_n+\mu_{n, \lambda }\right)-n_F\left(M_n-\mu_{n, \lambda }\right)\right] \right\}
=\lambda \epsilon_0 \mathcal{E},
\label{gapEq-Coulomb-DOS}
\end{equation}
where we used the definition for the surface charge densities in terms of the fermion propagator,
i.e., $\rho_{\lambda }=e\,\mbox{tr}[G_{\lambda}(u,u) \gamma^0]$.
Because the surface charge density is fixed by the external electric field, the electrochemical potential
$\mu^{(0)}_{\lambda}$ is not an independent parameter. It is determined together with the other dynamical
parameters by solving the system of Eqs.~(\ref{gapEq-Coulomb-gap-Delta}) through (\ref{gapEq-Coulomb-DOS}).

\section{Numerical results}
\label{sec:gap-equation-Coulomb-results-kappa}

In this section, we present our numerical solutions of gap equations (\ref{gapEq-Coulomb-gap-Delta}) through
(\ref{gapEq-Coulomb-gap-mu}), together with the constraint in Eq.~(\ref{gapEq-Coulomb-DOS}).
Before proceeding to the analysis, it is convenient to give the formal definition of the
Dirac and Haldane gaps in the TI model at hand. While the original surface gaps $m_{n,+}$
and $m_{n,-}$ (with $n \ge 1$) have a straightforward physical meaning, the symmetry
properties of the ground state can be better understood in terms of Dirac and Haldane gaps, i.e.,
\begin{equation}
m_{n, D}= \frac{m_{n,+}-m_{n,-}}{2}, \qquad m_{n, H}= \frac{m_{n,+}+m_{n,-}}{2}.
\label{gap-Coulomb-results-Dirac-Haldanem-tmu}
\end{equation}
Strictly speaking, these gaps cannot be associated with the usual Dirac and Haldane masses
in (2+1)-dimensional QED (see, Ref.~\cite{Appelquist:1986}), because $m_{\pm}$ in TIs correspond
to spatially separated surfaces. Since the free Hamiltonian (\ref{model-H-s-0}) contains the bare
Haldane gap $m^{(0)}$ due to the Zeeman interaction, it is also convenient to define
the dynamical part of the total Haldane gap $\Delta m_{n, H}\equiv m_{n,H}-m^{(0)}$.

In order to provide an insight into relation (\ref{gap-Coulomb-results-Dirac-Haldanem-tmu}) between
$m_{\pm}$ and Dirac and Haldane gaps, let us recall the reducible 4$\times$4 representation for
$\mbox{QED}_{2+1}$ considered in Ref.~\cite{Appelquist:1986}
\begin{eqnarray}
\tilde{\gamma}^0=\left(
                   \begin{array}{cc}
                     \gamma^0 & 0 \\
                     0 & -\gamma^0 \\
                   \end{array}
                 \right), \quad \tilde{\gamma}^1=\left(
                   \begin{array}{cc}
                     \gamma^1 & 0 \\
                     0 & -\gamma^1 \\
                   \end{array}
                 \right), \quad \tilde{\gamma}^2=\left(
                   \begin{array}{cc}
                     \gamma^2 & 0 \\
                     0 & -\gamma^2 \\
                   \end{array}
                 \right).
\label{gap-Coulomb-results-reducible}
\end{eqnarray}
In this representation, in addition to the $\gamma$-matrices in Eq.~(\ref{gap-Coulomb-results-reducible}),
there exist two other matrices,
\begin{eqnarray}
\tilde{\gamma}^3=i\left(
                   \begin{array}{cc}
                     0 & 1 \\
                     1 & 0 \\
                   \end{array}
                 \right), \quad \tilde{\gamma}^5=i\left(
                   \begin{array}{cc}
                     0 & 1 \\
                     -1 & 0 \\
                   \end{array}
                 \right),
\label{gap-Coulomb-results-reducible-35}
\end{eqnarray}
which anticommute with $\tilde{\gamma}^0$, $\tilde{\gamma}^1$, and $\tilde{\gamma}^2$. In terms of these
$4\times 4$ matrices, the existence of a $U(2)$ symmetry in the model of a TI slab,
defined by Eqs.~(\ref{model-H-0-matrices}) and (\ref{model-H-int}), is transparent.
The corresponding group generators are given by
\begin{equation}
1, \quad i{\cal R}_{\mu}\tilde{\gamma}^3, \quad {\cal R}_{\mu}\tilde{\gamma}^5, \quad \mbox{and}
\quad \tilde{\gamma}^3\tilde{\gamma}^5,
\label{gap-Coulomb-results-generators}
\end{equation}
where ${\cal R}_{\mu}$ is the operator which interchanges $\mu_{+}^{(0)}\leftrightarrow\mu_{-}^{(0)}$ in
the low-energy free Hamiltonian (\ref{model-H-0-matrices}). (Note that, in the absence of an external
electric field, there is no need in the operator ${\cal R}_{\mu}$.) As is easy to check, the Dirac gap
$m_D\bar{\Psi}\Psi$ breaks the $U(2)$ symmetry down to $U_{+}(1)\times U_{-}(1)$, where
$\bar{\Psi}=\Psi^{\dagger}\tilde{\gamma}^0$ and the subscript $\lambda=\pm$ labels the two
irreducible representations or the surfaces of the TI slab. The Haldane gap
$m_H\bar{\Psi}\tilde{\gamma}^3\tilde{\gamma}^5\Psi$ is invariant with respect to the $U(2)$
symmetry, but, unlike the Dirac gap, it breaks the parity $P$ and $T$ symmetries.
Since external electric and magnetic fields break $P$ and $T$ symmetries, the generation of the
Haldane gap has no effect on symmetry breaking. Therefore, only the dynamically generated
Dirac gap will spontaneously break the symmetry of our model. As we will see
below, such a gap is indeed generated due to the electrified magnetic catalysis.

\begin{figure*}[!t]
\begin{center}
\includegraphics[width=0.32\linewidth]{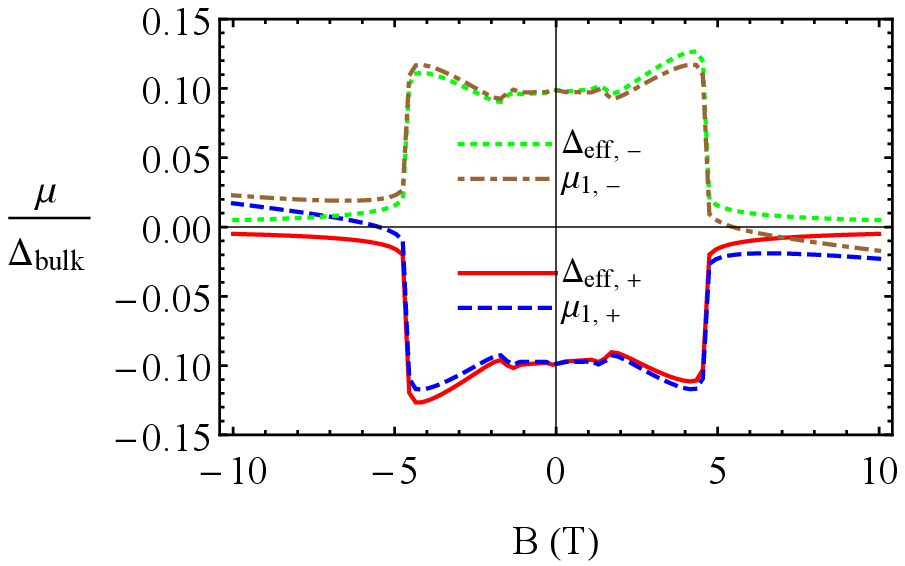}\hspace{0.012\linewidth}
\includegraphics[width=0.32\linewidth]{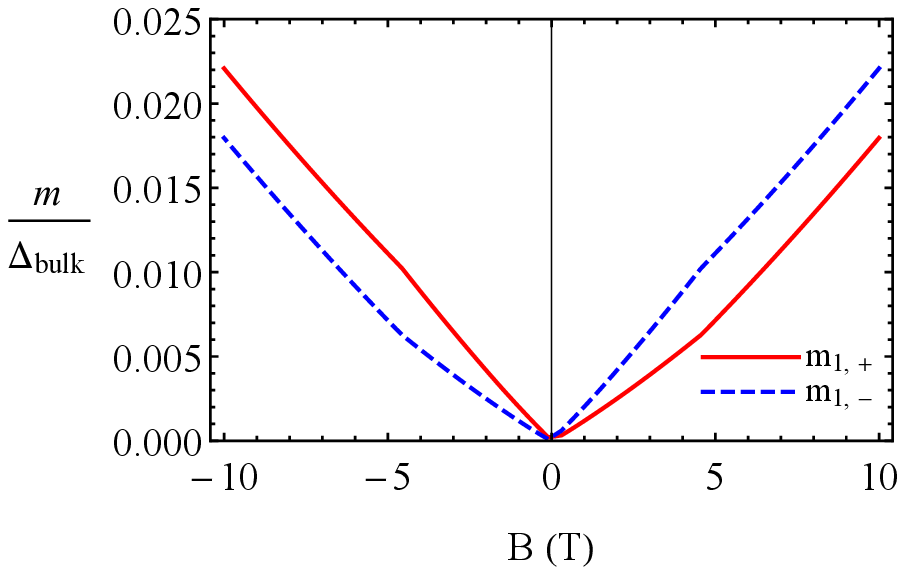}\hspace{0.012\linewidth}
\includegraphics[width=0.32\linewidth]{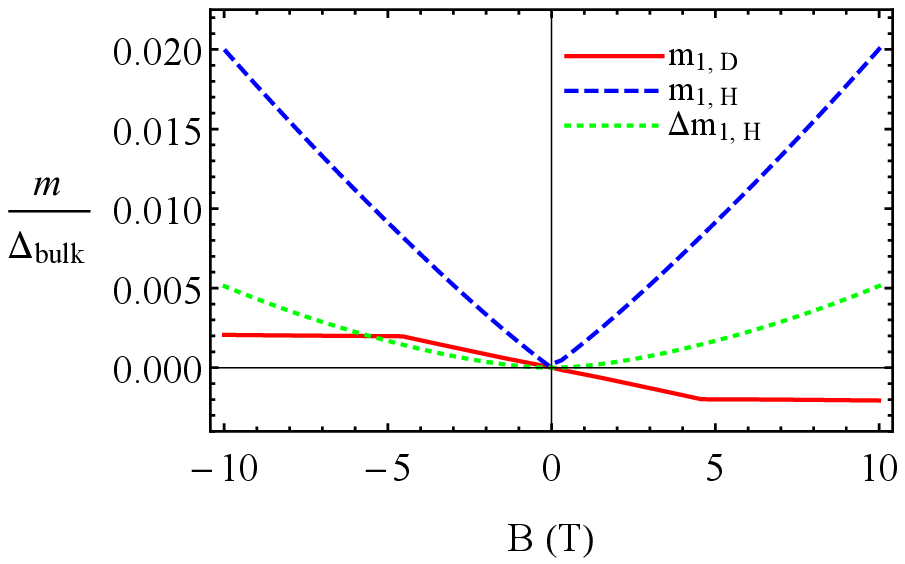}
\end{center}
\caption{(Color online) The lowest and first Landau level parameters as functions of the magnetic field
for fixed values of the external electric field, $\mathcal{E}=1~\mbox{mV/\AA}$, and temperature,
$T=5\times10^{-3}\Delta_{\rm bulk}\approx20~\mbox{K}$. The results for the (effective) electrochemical
potentials $\Delta_{\mathrm{eff}, \pm}$, and $\mu_{1,\pm}$ are shown in the left panel, the gaps $m_{1,\pm}$
are shown in the middle panel, and Dirac and Haldane gaps $m_{1,D}$, $m_{1,H}$, and $\Delta m_{1,H}$
are shown in the right panel.}
\label{fig:gapEq-Coulomb-results-surface-tmu0-HLL-B-kappa}
\end{figure*}

For numerical calculations, it is useful to estimate energy scales in the problem at hand
\begin{eqnarray}
\Delta_{\rm bulk}\approx 350~\mbox{meV},\quad\quad \frac{g_s\mu_B B}{2}\approx 0.5B[T]~\mbox{meV},\nonumber \\
\epsilon_B=\sqrt{2v_F^2|eB|}\approx 22.6\sqrt{B[T]}~\mbox{meV},\quad\quad l\approx25.7~\mbox{nm}/\sqrt{B[T]}.
\label{gap-Coulomb-results-energy-scales}
\end{eqnarray}
By solving numerically the gap equations (\ref{gapEq-Coulomb-gap-Delta}), (\ref{gapEq-Coulomb-gap-m})
and (\ref{gapEq-Coulomb-gap-mu}) together with constraint (\ref{gapEq-Coulomb-DOS}), we straightforwardly
obtain the electrochemical potentials $\mu_{n,\pm}$ and the gaps $m_{n,\pm}$ as functions of the magnetic field.
The results for the lowest and first Landau level parameters are shown in
Fig.~\ref{fig:gapEq-Coulomb-results-surface-tmu0-HLL-B-kappa} for fixed values
of the electric field and temperature, $\mathcal{E}=1~\mbox{mV/\AA}$ and $T=5\times10^{-3}\Delta_{\rm bulk}
\approx20~\mbox{K}$, respectively. In the calculation, we truncated the system of equations by including only
$n_{\rm max}=26$ Landau levels.

%%%%%%%%%%%%%%%%%%
%%%%%%%%%%%%%%%%%%

As we see from the left panel in Fig.~\ref{fig:gapEq-Coulomb-results-surface-tmu0-HLL-B-kappa}, the absolute
values of the electrochemical potentials $\Delta_{\mathrm{eff}, \pm}$ and $\mu_{1,\pm}$ experience a large
jump around $|B|\approx5~\mbox{T}$. The jump corresponds to the point at which the filling of the first Landau
level starts. We checked that the position of the jump shifts to larger values of the magnetic field with increasing
the external electric field. Of course, this behavior is expected, since larger electric fields require higher charge
densities on the TI surfaces. In addition to the large jump around $|B|\approx5~\mbox{T}$, we also
observe additional features in the dependence of $\Delta_{\mathrm{eff}, \pm}$ and $\mu_{1,\pm}$
at smaller values of the magnetic field. They generically correspond to the onset of filling of higher Landau levels.
Here it is appropriate to mention that, in all regimes studied, the electrochemical potentials $\mu^{(0)}_{\pm}$
are very similar quantitatively to $\mu_{1,\pm}$ and, therefore, we do not show them in our figures.

\begin{figure*}[!t]
\begin{center}
\includegraphics[width=0.32\linewidth]{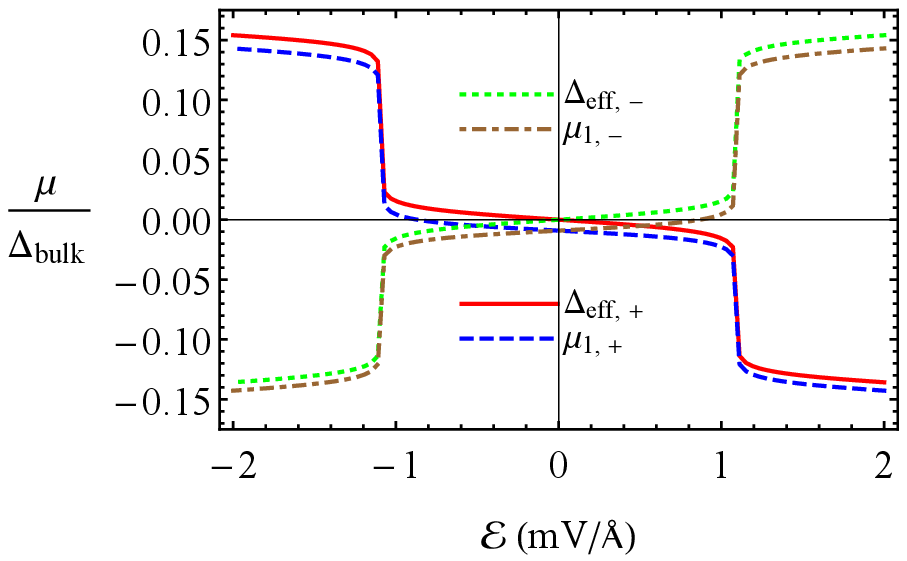}\hspace{0.012\linewidth}
\includegraphics[width=0.32\linewidth]{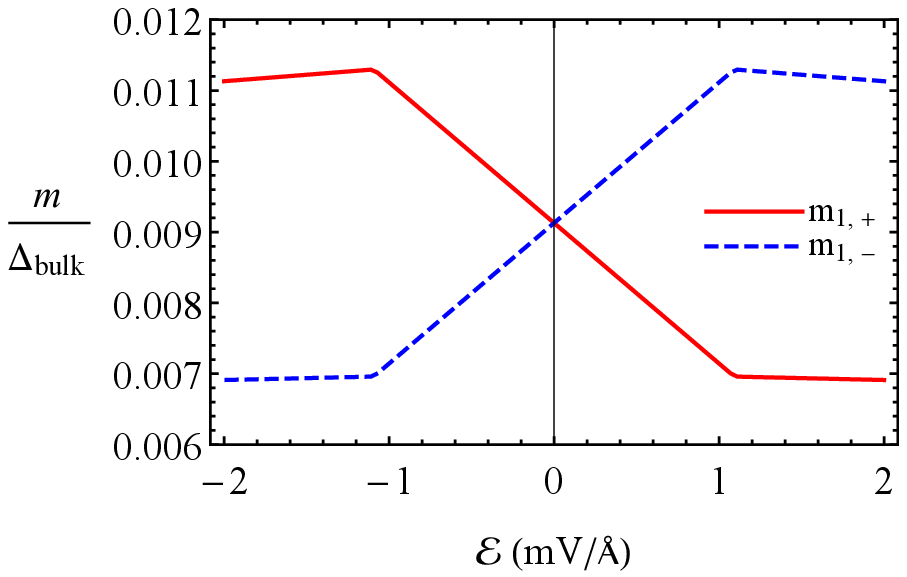}\hspace{0.012\linewidth}
\includegraphics[width=0.32\linewidth]{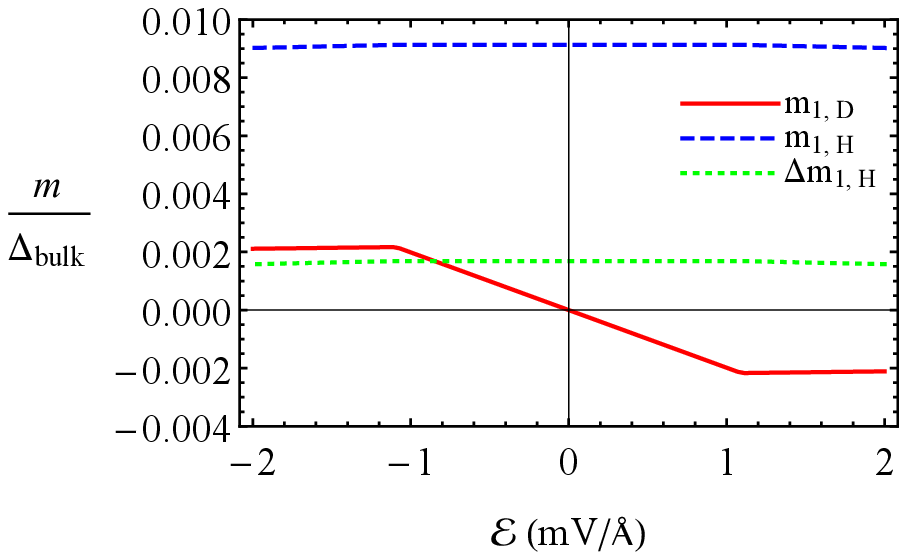}
\end{center}
\caption{(Color online) The lowest and first Landau level parameters as functions of the external electric field for fixed
values of the magnetic field, $B=5~\mbox{T}$, and temperature, $T=5\times10^{-3}\Delta_{\rm bulk}\approx20~\mbox{K}$.
The results for the (effective) electrochemical potentials $\Delta_{\mathrm{eff}, \pm}$, and $\mu_{1,\pm}$ are
shown in the left panel, the gaps $m_{1,\pm}$ are shown in the middle panel, and Dirac and Haldane gaps
$m_{1,D}$, $m_{1,H}$ and $\Delta m_{1,H}$ are shown in the right panel.}
\label{fig:gapEq-Coulomb-results-surface-tmu0-HLL-E-kappa}
\end{figure*}

Let us now turn to the discussion of the dynamically generated gaps. The results in the middle panel of
Fig.~\ref{fig:gapEq-Coulomb-results-surface-tmu0-HLL-B-kappa} clearly demonstrate that the surface
gaps $m_{1,\pm}$ monotonically increase with the magnetic field. More interestingly, however, we find that the
values of the gaps on the two surfaces, $m_{1,+}$ and $m_{1,-}$, remain comparable, although not identical
to each other for sufficiently weak electric fields. The importance of this observation becomes obvious
in the context of the $U(2)$ symmetry discussed earlier. Indeed, if the values of $m_{1,+}$ and
$m_{1,-}$ were exactly the same, they would describe a pure Haldane solution. As is clear
from the definition in Eq.~(\ref{gap-Coulomb-results-Dirac-Haldanem-tmu}), a small difference
between $m_{1,+}$ and $m_{1,-}$ implies the existence of a dynamically generated Dirac gap.
Such a gap is induced by the applied electric field. This conclusion is further supported by
the dependence of the dynamical gaps on the electric field, shown in
Fig.~\ref{fig:gapEq-Coulomb-results-surface-tmu0-HLL-E-kappa} and discussed below Eq.~(\ref{gapEq-results-Coulomb-DOS}).

The results in the right panel of Fig.~\ref{fig:gapEq-Coulomb-results-surface-tmu0-HLL-B-kappa}
demonstrate that the absolute value of the Dirac gap $m_{1, D}$ increases with the magnetic field at
sufficiently small fields, $|B|\lesssim 5~\mbox{T}$, when the LLL is fully filled. At larger
magnetic fields, $|B|\gtrsim5~\mbox{T}$, when the LLL is not fully filled, the Dirac gap remains
nearly constant (or increases very slowly). In contrast, the dynamical part of the Haldane gap
$\Delta m_{1, H}$ increases approximately as $B^2$. Note that, because of the linear dependence
of $m^{(0)}$ on the magnetic field, the total Haldane gap $m_{1, H}$ grows almost linearly with $B$.

%%%%%%%%%%%%%%%%%%
%%%%%%%%%%%%%%%%%%

The dependencies of the lowest and first Landau level parameters on the external electric field are presented
in Fig.~\ref{fig:gapEq-Coulomb-results-surface-tmu0-HLL-E-kappa} for fixed values of the magnetic field
($B=5~\mbox{T}$) and temperature ($T=5\times10^{-3}\Delta_{\rm bulk}\approx20~\mbox{K}$). As we see from
the left panel in Fig.~\ref{fig:gapEq-Coulomb-results-surface-tmu0-HLL-E-kappa}, the absolute values of the
electrochemical potentials slowly increase with electric field at first, and then experience a substantial jump at
$|\mathcal{E}|\approx1~\mbox{mV/\AA}$. The jump corresponds to the field at which the filling of the first Landau
level begins. From the middle panel in Fig.~\ref{fig:gapEq-Coulomb-results-surface-tmu0-HLL-E-kappa}, we see
that the surface gaps $m_{1,\pm}$ have a linear dependence at weak fields (i.e., in the regime of a partially
filled LLL) and stay approximately constant at higher electric fields. As might have been expected, the
Haldane gap $m_{1,H}$, which is shown in the right panel of Fig.~\ref{fig:gapEq-Coulomb-results-surface-tmu0-HLL-E-kappa},
depends very weakly on the applied electric field. This is in contrast to the behavior of the Dirac gap $m_{1, D}$
(see the right panel in Fig.~\ref{fig:gapEq-Coulomb-results-surface-tmu0-HLL-E-kappa}), which is linear in $\mathcal{E}$
at small fields and stays approximately constant at large fields. It should be also emphasized that the Dirac gap
vanishes at $\mathcal{E}=0$. As we argue below, this fact is important from the viewpoint of symmetry properties
in the model.

As already suggested earlier, the generation of the Dirac gap is directly connected with the applied external electric
field. In order to demonstrate this in the simplest possible setting, it is instructive to consider an approximate form of the gap
equation (\ref{gapEq-Coulomb-gap-m}) in the limit of a large magnetic field. By rewriting it in terms of the Haldane and
Dirac gaps, we obtain
\begin{eqnarray}
m_{n, H}&\approx &m^{(0)}+\frac{1}{4l} \sum_{n^{\prime}=1}^{\infty} \left(\alpha v_F \mathcal{K}^{(0)}_{n^{\prime}-1,n-1}
+  \alpha v_F \mathcal{K}^{(0)}_{n^{\prime},n} +\frac{G_{\rm int}}{\pi l} \right) \frac{m_{n^{\prime},H} }{M_{n^{\prime}}} ,
\label{gapEq-results-Coulomb-gap-mH} \\
m_{n, D}&\approx& -\pi l s_B  \epsilon_0\left(\alpha  v_F \mathcal{K}^{(0)}_{0,n}+\frac{G_{\rm int}}{2\pi l}\right)\frac{\mathcal{E}}{e}
+ \frac{1}{4l} \sum_{n^{\prime}=1}^{\infty} \left(\alpha v_F \mathcal{K}^{(0)}_{n^{\prime}-1,n-1}
+  \alpha v_F \mathcal{K}^{(0)}_{n^{\prime},n} +\frac{G_{\rm int}}{\pi l} \right) \frac{m_{n^{\prime},D} }{M_{n^{\prime}}} ,
\label{gapEq-results-Coulomb-gap-mD}
\end{eqnarray}
where we took into account that $n_F(M_{n}\pm \mu_{n, \lambda }) \ll 1$ for $n\geq 1$ and assumed that
$M_{n^{\prime}}$ is almost independent of the small Dirac gap. Note that in order to rewrite the LLL contributions in the gap equations in terms of the electric field $\mathcal{E}$, we used the following approximate expression for the surface charge densities:
\begin{equation}
\rho_{+}= -\rho_{-}=\frac{\rho_{+}-\rho_{-}}{2}\approx\frac{e}{4\pi l^2} \left[n_F\left(\Delta_{\mathrm{eff}, +}\right) -n_F\left(\Delta_{\mathrm{eff}, -}\right) \right] =\epsilon_0\mathcal{E}.
\label{gapEq-results-Coulomb-DOS}
\end{equation}
By comparing the gap equations (\ref{gapEq-results-Coulomb-gap-mH}) and (\ref{gapEq-results-Coulomb-gap-mD}),
we see that the external electric field plays the role of a ``seed" for the Dirac gap $m_{n, D}$, just as the bare
gap parameter $m^{(0)}$ for the Haldane gap $m_{n,H}$. This explains why the external electric field is the key
factor in generating the Dirac gap and breaking the $U(2)$ symmetry in the slab of 3D TIs.

\begin{figure*}[!t]
\begin{center}
\includegraphics[width=0.32\linewidth]{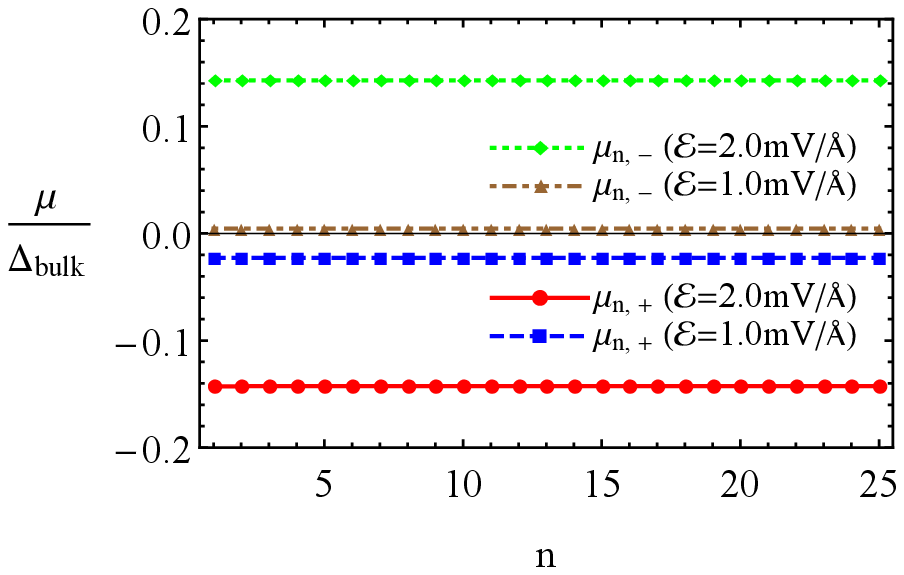}\hspace{0.012\linewidth}
\includegraphics[width=0.32\linewidth]{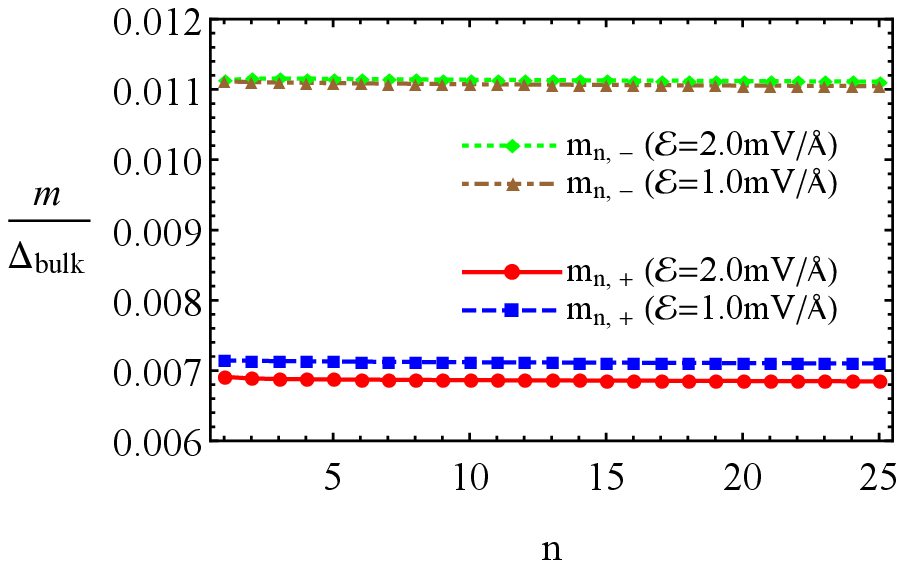}\hspace{0.012\linewidth}
\includegraphics[width=0.32\linewidth]{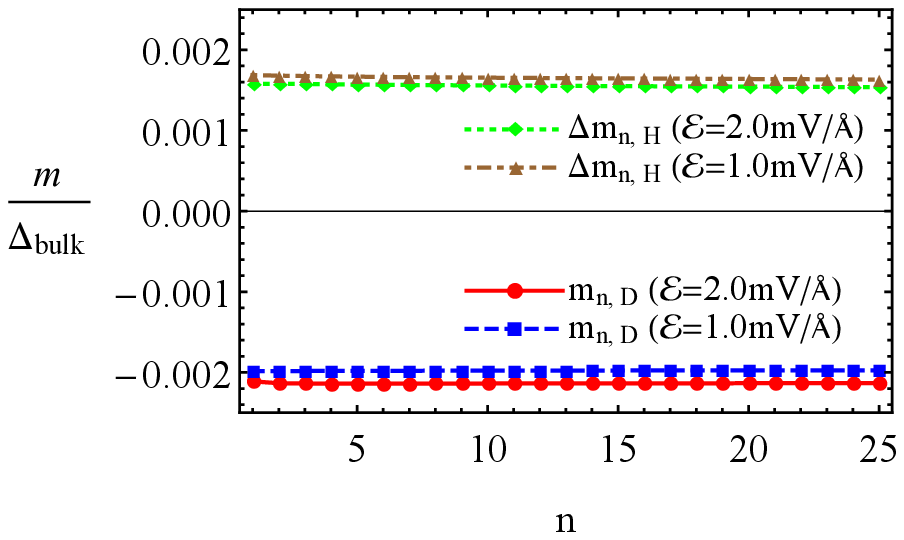}
\end{center}
\caption{(Color online) The electrochemical potentials $\mu_{n,\pm}$ (left panel), the gaps $m_{n,\pm}$
(middle panel), and the Dirac and Haldane gaps (right panel) as functions of the Landau level index
$n$. The results for $\mathcal{E}=2~\mbox{mV/\AA}$ are represented by red solid and green
dotted lines. The results for $\mathcal{E}=1~\mbox{mV/\AA}$ are represented by blue dashed
and brown dash-dotted lines. The values of the magnetic field and temperature are $B=5~\mbox{T}$
and $T=5\times10^{-3}\Delta_{\rm bulk}\approx20~\mbox{K}$, respectively.}
\label{fig:gapEq-Coulomb-results-surface-tmu0-HLL-E-n-kappa}
\end{figure*}

It may be instructive to study the dependence of the electrochemical potentials $\mu_{n,\pm}$ and gaps
$m_{n,\pm}$ on the Landau level index $n$. The corresponding results for two different values of the
electric field are presented in Fig.~\ref{fig:gapEq-Coulomb-results-surface-tmu0-HLL-E-n-kappa}
for $B=5~\mbox{T}$ and $T=5\times10^{-3}\Delta_{\rm bulk}\approx20~\mbox{K}$. As we see, all
dynamical parameters depend very weakly on the Landau level index $n$.
In view of the large surface dielectric constant and, consequently, weak Coulomb interaction, this result is not surprising. Moreover, it strongly suggests that the long-range interaction indeed plays a minor role compared to the local interaction.

%%%%%%%%%%%%%%%%%%
%%%%%%%%%%%%%%%%%%

\begin{figure*}[!t]
\begin{center}
\includegraphics[width=0.42\linewidth]{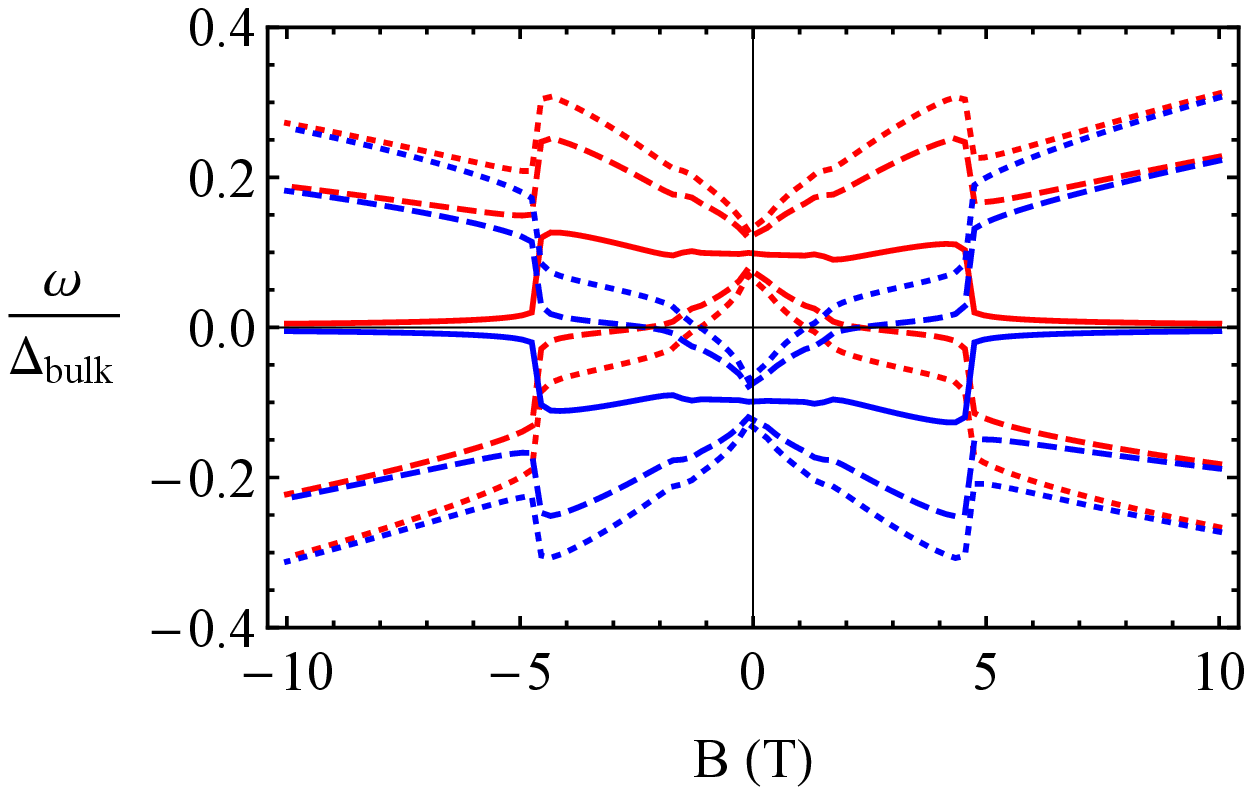}\hspace*{0.1\linewidth}
\includegraphics[width=0.42\linewidth]{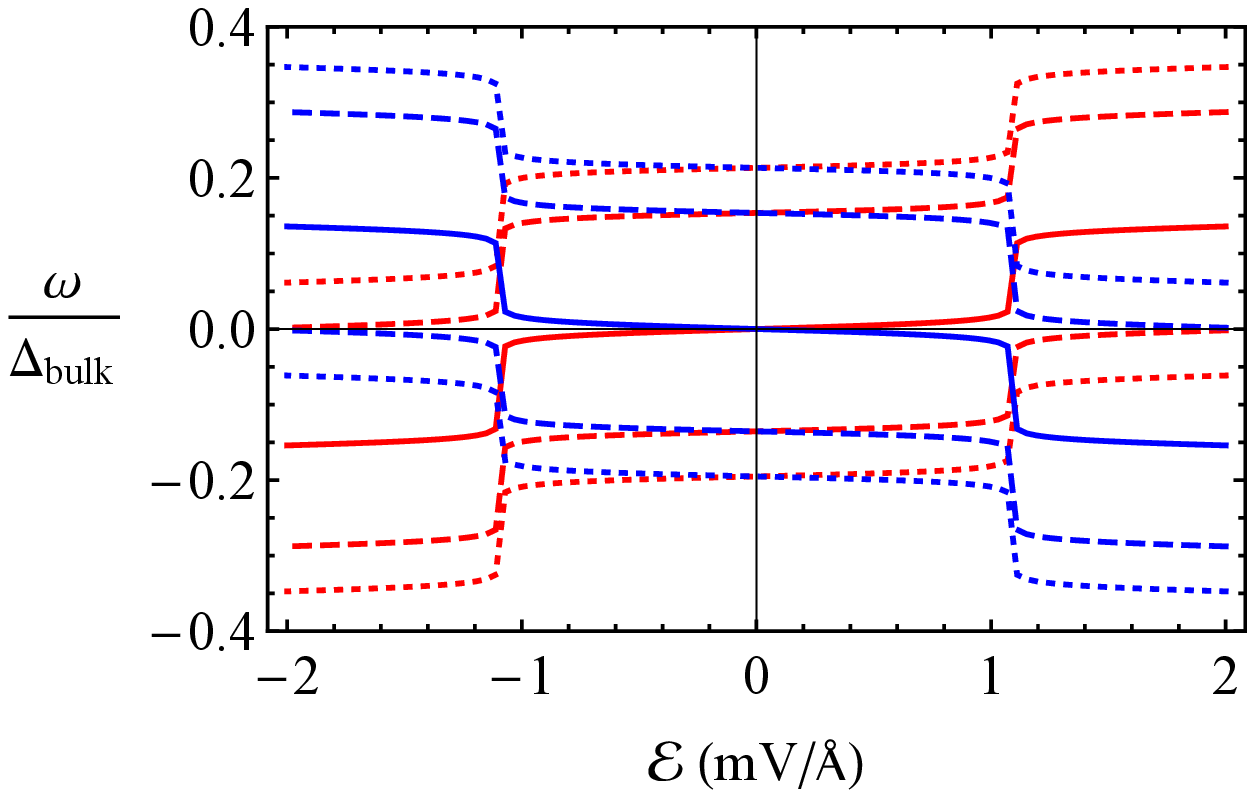}
\end{center}
\caption{(Color online) The quasiparticle energies for the first three Landau levels as functions of the
magnetic field at fixed $\mathcal{E}=1~\mbox{mV/\AA}$ (left panel) and as functions of the electric
field at fixed $B=5~\mbox{T}$ (right panel). Red and blue lines denote the quasiparticle
energies on the top and bottom surfaces, respectively. Solid lines represent the LLL, dashed and dotted lines correspond to the first and second Landau levels, respectively. The temperature is
$T=5\times10^{-3}\Delta_{\rm bulk}\approx20~\mbox{K}$.}
\label{fig:gapEq-Coulomb-results-surface-spectrum}
\end{figure*}

By using the above results, we can also obtain the quasiparticle energy levels as functions of the magnetic and electric fields
\begin{eqnarray}
\omega_{0,\lambda}=-\Delta_{\rm eff, \lambda}, \quad \omega_{n>0, \lambda}=-\mu_{n, \lambda}\pm M_{n},
\label{gap-Coulomb-results-spectrum}
\end{eqnarray}
where $M_n$ were given below Eq.~(\ref{gapEq-Coulomb-G-no-phase}). The corresponding numerical results are summarized in
Fig.~\ref{fig:gapEq-Coulomb-results-surface-spectrum}.

As we see from the left panel in Fig.~\ref{fig:gapEq-Coulomb-results-surface-spectrum}, there is a rather
large splitting between the energy levels on the top and bottom surfaces at small values of the magnetic field.
This corresponds to the regime with higher Landau levels being occupied. With increasing the magnetic field,
the magnitude of splitting quickly diminishes and becomes rather small when the LLL regime is reached. In
contrast, the increase of the electric field tends to amplify the splitting between the Landau levels.
The existence of such a splitting may lead to an observation of new plateaus in the Hall conductivity. The large
jumps in the energy spectrum at $|B|\approx5~\mbox{T}$ and $|\mathcal{E}|\approx1~\mbox{mV/\AA}$
correspond to the onset of and the exit from the LLL regime, respectively. As is clear, these features are
directly connected with the corresponding jumps in the electrochemical potentials, seen in the left panels
of Figs.~\ref{fig:gapEq-Coulomb-results-surface-tmu0-HLL-B-kappa} and
\ref{fig:gapEq-Coulomb-results-surface-tmu0-HLL-E-kappa}.

%%%%%%%%%%%%%%%%%%
%%%%%%%%%%%%%%%%%%

Before concluding this section, let us briefly discuss the role of finite temperature in our solution.
As expected, the main results remain qualitatively the same for a whole range
of sufficiently small values of the temperature. With increasing (decreasing) the temperature, however,
the jumps that correspond to the onset of and the exit from the LLL regime become smoother (sharper)
in the dependence of the electrochemical potentials on the fields, shown in the left panels of
Figs.~\ref{fig:gapEq-Coulomb-results-surface-tmu0-HLL-B-kappa} and
\ref{fig:gapEq-Coulomb-results-surface-tmu0-HLL-E-kappa}. It is also worth pointing that a weak
dependence of electrochemical potentials on the fields in the regions between the jumps is caused by
thermal broadening of Landau levels. It vanishes in the limit $T\to0$.

\section{Inhomogeneous phase with two stripes: qualitative approach}
\label{sec:stripe}

In the previous section, we advocated the homogeneous phase with dynamically generated gaps as the ground
state of 3D TIs in a sufficiently strong external electric field. On the other hand, the inhomogeneous CDW phase
considered in Ref.~\cite{Vishwanath:2010} is likely to be more favorable in weak electric fields.
In order to provide a qualitative analytic description of the inhomogeneous CDW phase with a ``stripe'' pattern,
in this section we consider a simple configuration of two stripes with an infinitely thin transition region, or a
domain wall at $x=0$. This is modeled by an inhomogeneous gap $m(x)=|m|\sign{(x)}$ for the top surface, i.e., $\lambda=+1$ and
$s_{B}=+1$. (Note that such a gap function with asymptotes of opposite sign at $x\to\pm\infty$, but without a
magnetic field, is qualitatively similar to the famous Jackiw-Rebbi solution in 1D \cite{Jackiw-Rebbi}.)

\begin{figure*}[!t]
\begin{center}
\includegraphics[width=0.32\linewidth]{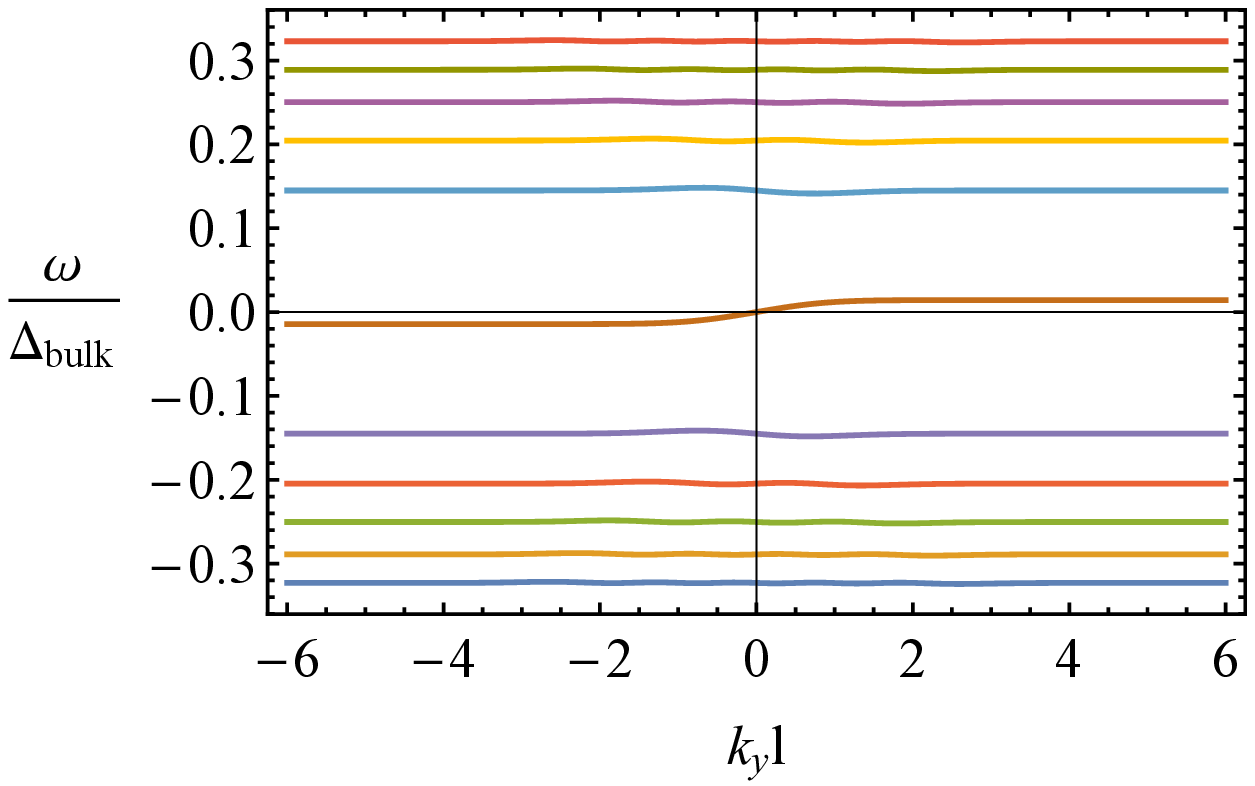}\hspace{0.012\linewidth}
\includegraphics[width=0.32\linewidth]
{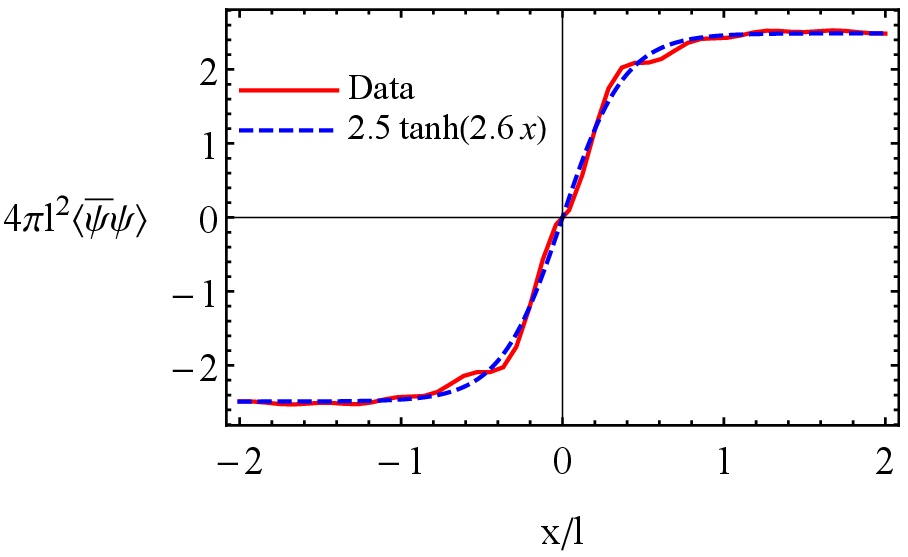}\hspace{0.012\linewidth}
\includegraphics[width=0.32\linewidth]
{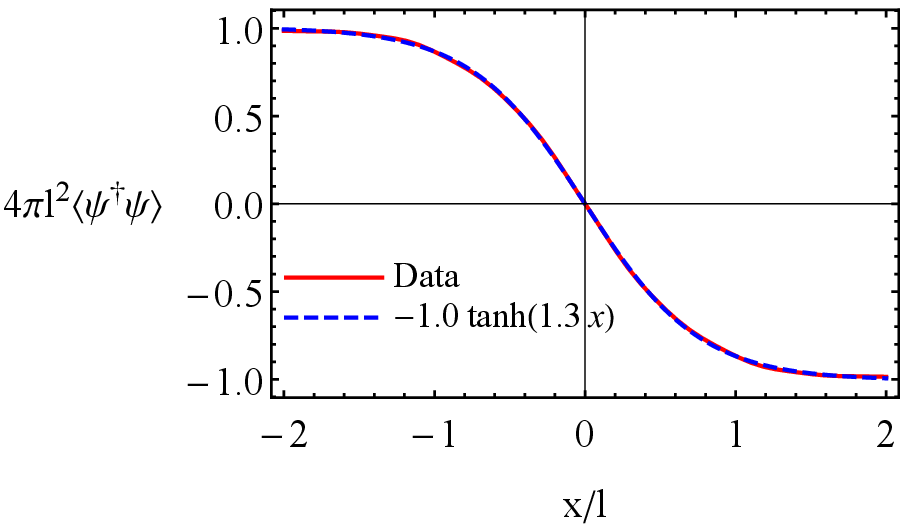}
\end{center}
\caption{(Color online) The energy spectrum as a function of $k_y$ (left panel), the chiral condensate (middle panel) and
the charge density (right panel) as functions of $x$ in the Dirac problem with the inhomogeneous gap $m(x)=|m|\sign{(x)}$.
For simplicity, only the first 6 Landau levels are presented in the left panel. The red solid lines in the middle and right panels
represent the numerical data. The blue dashed lines denote the corresponding fits. The model parameters are
$m=5~\mbox{meV}$, $\mu=0$, and $B=5~\mbox{T}$.}
\label{fig:stripe-zero-modes-chiral-condensate}
\end{figure*}

The solution to the Dirac equation with the gap function in the form $m(x)=|m|\sign{(x)}$ is discussed in
Appendix~\ref{sec:domain-wall}. The corresponding numerical results for the quasiparticle energy spectrum
as a function of $k_y$, as well as the chiral condensate and charge density as functions of the spatial coordinate
$x$, are shown in Fig.~\ref{fig:stripe-zero-modes-chiral-condensate}. In order to plot the results, we fixed the
model parameters as follows: $m=5~\mbox{meV}$, $\mu=0$, and $B=5~\mbox{T}$. In the calculation,
we also limited the sum over Landau levels ($n_{\rm max}=26$) and cut off the integration over $k_y$
($-6/l \leq k_y \leq 6/l$).

%%%%%%%%%%%%%%%%%%
%%%%%%%%%%%%%%%%%%

As we see from Fig.~\ref{fig:stripe-zero-modes-chiral-condensate}, the chiral condensate and the charge density
have a kink and antikink structure, respectively. Therefore, the existence of zero energy states on the domain wall agrees with the inhomogeneous form of the gap function. The chiral condensate and the charge density
can be fitted well by the following functions:
\begin{eqnarray}
\label{stripe-zero-modes-chircond-fit}
\mbox{tr}[G(u,u)]=\sum_{n} \frac{\sign{(\omega_n)}}{2} \bar{\psi}_{\omega_n}(x)\psi_{\omega_n}(x)
\approx \frac{1}{4\pi l^2} 2.5 \tanh{\left(2.6\frac{x}{l}\right)},
\\
\label{stripe-zero-modes-DOS-fit}
\mbox{tr}[\gamma^0 G(u,u)]=\sum_{n} \frac{\sign{(\omega_n)}}{2} \psi^{\dag}_{\omega_n}(x)\psi_{\omega_n}(x)
\approx -\frac{1}{4\pi l^2} 1.0 \tanh{\left(1.3\frac{x}{l}\right)},
\end{eqnarray}
where the sum runs over the complete set of eigenstates, given by the solutions to the spectral equation
(\ref{stripe-zero-modes-spectral-eq}). These fits are shown in Fig.~\ref{fig:stripe-zero-modes-chiral-condensate}
alongside with the numerical solutions.

Let us now consider the case of a nonzero external electric field, applied perpendicularly to the slab. By considering
a sufficiently thick slab, we will assume that the average electric field inside the slab vanishes. In the homogeneous
case, the field is screened by the uniform surface charge densities. This generically requires specific nonzero
electrochemical potentials $\mu_{\pm}$ on the top and bottom TIs surfaces. In the inhomogeneous striped
phase, however, the simplest way to achieve a nonzero average surface charge densities is to vary the width of stripes
by $\Delta l_x$. The value of $\Delta l_x$ can be estimated from the following expression:
\begin{equation}
\epsilon_0\mathcal{E} = \rho(l_x+\Delta l_x) - \rho(l_x)+ \rho(-l_x+\Delta l_x) - \rho(-l_x)
\approx \frac{e}{4\pi l^2}
\left[\tanh{\left(1.3\frac{l_x+\Delta l_x}{l}\right)}-\tanh{\left(1.3\frac{l_x-\Delta l_x}{l}\right)}\right].
\label{stripe-energy-diff-stripes-DOS}
\end{equation}
where we used the standard definition for the surface charge density $\rho=e\,\mbox{tr}[\gamma^0 G(u,u)]$ together with Eq.~(\ref{stripe-zero-modes-DOS-fit}).

%%%%%%%%%%%%%%%%%%
\begin{figure*}[!t]
\begin{center}
\includegraphics[width=0.45\linewidth]{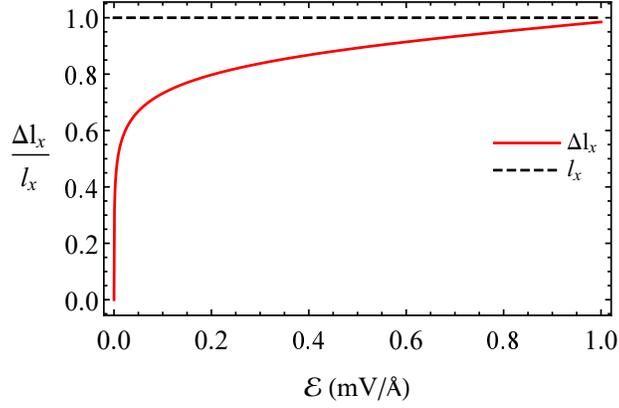}
\end{center}
\caption{(Color online) The ratio of the width correction $\Delta l_x$ to the stripe half-width $l_x$ as a function
of the external electric field (red  line). The model parameters are $m=5~\mbox{meV}$, $\mu=0$,
$B=5~\mbox{T}$, and $l_x=500~\mbox{\AA}$.}
\label{fig:stripe-energy-diff-nonzero-field-Lky-kappa}
\end{figure*}
%%%%%%%%%%%%%%%%%%

Our numerical result for the ratio of the correction $\Delta l_x$ to the stripe half-width $l_x$ is
plotted in Fig.~\ref{fig:stripe-energy-diff-nonzero-field-Lky-kappa}. We see that the correction to
the stripe width $\Delta l_x$ becomes significant at sufficiently strong electric fields. In fact, it is
comparable to $l_x$ already for $\mathcal{E} \gtrsim 1~\mbox{mV/\AA}$. This suggests that the
stripe phase is unstable when the electric field exceeds a certain critical value $\mathcal{E}_{\rm cr}$.
Quantitatively, the critical value roughly corresponds to the beginning of the first Landau level filling,
\begin{equation}
\mathcal{E}_{\rm cr}\approx\frac{e^2B}{4\pi \epsilon_{0} \hbar c},
\label{stripe-energy-diff-critical-E}
\end{equation}
where we used Eq.~(\ref{gapEq-Coulomb-DOS}) and restored Plank's constant $\hbar$ and the
speed of light $c$. This relation implies that $\mathcal{E}_{\rm cr}\approx0.22B[\mbox{T}]~\mbox{mV}/\mbox{\AA}$,
which is in a good agreement with our previous numerical estimate $\mathcal{E} \gtrsim 1~\mbox{mV/\AA}$ at $B=5~\mbox{T}$.

Another way to estimate the critical value of the electric field is to compare the free energy density
in the homogeneous phase, which is given by Eq.~(\ref{app-free-energy-T}), with the energy density
of the stripe phase estimated in Ref.~\cite{Vishwanath:2010}. In other words, the value of
$\mathcal{E}_{\rm cr}$ is given by the solution to the following equation:
\begin{equation}
0=\sum_{\lambda}\Omega_{\lambda} - \left[\frac{\sqrt{\alpha \gamma}}{l^2}-\frac{m}{l^2}\right],
\label{stripe-energy-diff-critical-field-Eq}
\end{equation}
where the term in square brackets corresponds to the energy cost of creating the stripe phase
\cite{Vishwanath:2010}. The latter is characterized by the domain wall tension $\gamma\sim1/l^2$
and the magnetic mass $m\sim 1/l$. The value of $\alpha$ is given below Eq.~(\ref{gapEq-Coulomb-gap-1}).
The solution to Eq.~(\ref{stripe-energy-diff-critical-field-Eq}) can be easily obtained numerically and appears
to agree quite well with the estimate in Eq.~(\ref{stripe-energy-diff-critical-E}). One can also obtain an
approximate analytical solution to Eq.~(\ref{stripe-energy-diff-critical-field-Eq}) by using the LLL
approximation for the free energy density (\ref{app-free-energy-T}), i.e.,
\begin{equation}
\Omega_{\lambda} \approx \frac{\lambda \epsilon_0 \mathcal{E} \Delta_{\rm eff}}{e} ,
\label{stripe-energy-diff-free-energy-LLL-1}
\end{equation}
where we used the LLL approximation for the charge density (\ref{gapEq-Coulomb-DOS}).
Then, by substituting this into Eq.~(\ref{stripe-energy-diff-critical-field-Eq}) and estimating
the effective electrochemical potential as $\Delta_{\rm eff}\sim \lambda/l$, we find the following
critical value of the electric field:
\begin{equation}
\mathcal{E}_{\rm cr} \approx \frac{e^2B (1-\sqrt{\alpha})}{2\epsilon_0 \hbar c}.
\label{stripe-energy-diff-critical-E-2}
\end{equation}
This result is qualitatively the same as the estimate in Eq.~(\ref{stripe-energy-diff-critical-E}),
although quantitatively appears to be somewhat larger, $\mathcal{E}_{\rm cr} \approx
1.03B[\mbox{T}]~\mbox{mV}/\mbox{\AA}$. We conclude, therefore, that the critical electric
field scales linearly with the magnetic field, $\mathcal{E}_{\rm cr} \sim e^2B/(\epsilon_0 \hbar c)$,
but the coefficient of proportionality is determined only up to an overall factor of order $1$.

\section{Discussion}
\label{sec:Discussion}

In this section, we discuss the range of validity and limitations of our study, and compare our
main results with those existing in the literature.

Let us start by pointing the limitation of our model used for the description of the TI surface states.
While the model captures the Dirac nature of the low-energy quasiparticles, it does not describe
the hexagonal warping of the Fermi surface that occurs away from the Dirac point \cite{Chen:2009, Hsieh:2009}.
The corresponding effect was taken into account in the study of gap generation in Ref.~\cite{Baum:2012}
and could play an essential role in some TIs. For example, this may be the case in Bi$_2$Te$_3$ [e.g.,
see Fig.~1(c) in Ref.~\cite{Chen:2009}], in which the band gap is about three times smaller than in
Bi$_2$Se$_3$ and the trigonal potential $\sim k^3$ is rather strong. In the case of Bi$_2$Se$_3$,
however, the hexagonal warping could be safely neglected, except for the case of rather high values of
the chemical potential [e.g., see Fig.~8 in Ref.~\cite{Ando:rev} and Fig.~3(b) in Ref.~\cite{Cava-Hasan}].
The model used in this study also ignores a Schr\"{o}dinger-type term $\sim k^2$, which
describes an asymmetry between the electron and hole bands \cite{Carbotte:2015} (see also Fig.~1 in
Ref.~\cite{Cava-Hasan}). When the quadratic term is sufficiently small, it is not expected to substantially
affect the dynamics of the gap generation.

The study here did not include the effects of the intersurface tunneling on the dynamical
generation of gaps. According to Ref.~\cite{Linder:2009}, tunneling between the opposite surfaces
may be quite important only for sufficiently thin ($l_z\lesssim8~\mbox{nm}$) TI slabs.
Therefore, neglecting the intersurface tunneling is expected to be a good approximation in the case
of thick samples. It would be interesting, however, to rigorously study the corresponding
effects in thin TI films in external electric and magnetic fields.

One of the uncertainties of the model Hamiltonian used in this study is the strength of the local interaction
$G_{\rm int}$. Although the order of magnitude of this coupling constant could be estimated by using general
arguments, its precise value is unknown. Despite this, we argue that the simplified model (\ref{model-H-0-matrices})
that includes both short- and long-range interactions (\ref{model-H-int}) is sufficient for the qualitative analysis
of the electrified magnetic catalysis in 3D TIs. Moreover, we might even suggest that, irrespective of the
specific value of the coupling constants, the qualitative features established here should be rather universal.

It is interesting to compare our results with those obtained in Ref.~\cite{Vishwanath:2010}, where the
phase diagram was studied in 3D TIs in a magnetic field, but without an external electric field. The
authors of Ref.~\cite{Vishwanath:2010} argued that, depending on the strength of local interaction,
the CFL or CDW (``stripe'' or ``bubble'')
phases can be realized. Our results here suggest that neither of those two phases describe the ground
state of the TI slab in a sufficiently strong external electric field. The CFL phase with the half-filled LLL
on each TI surface cannot be easily deformed to screen out the external electric field from penetrating
into the TI bulk. This would imply a large energy cost and disfavor the CFL phase. The CDW phase
could perhaps survive when a relatively weak electric field is applied. In this case, the average charge
densities on the TI surfaces, which are needed to screen the electric field out from the bulk, could be
simply obtained by the formation of positive and negative stripes (or bubbles) of unequal size. (Note
that it is energetically favorable to have either completely filled or empty LLL inside the stripes \cite{Vishwanath:2010}.)
As we showed in Sec.~\ref{sec:stripe}, a simple estimate suggests that the charge imbalance obtained
by the variation of the stripe widths can compensate only relatively weak electric fields. Therefore, a
sufficiently strong electric field $\mathcal{E}>\mathcal{E}_{\rm cr}$ also destroys the CDW phase.
Our parametric estimate for the critical electric field strength is $\mathcal{E}_{\rm cr} \sim e^2B/(\epsilon_0 \hbar c)$.

In view of the above arguments, we claim that the ground state of the TI slab in a nonzero magnetic field
and a sufficiently strong electric field is a homogeneous phase with equal in magnitude, but opposite in
sign surface charge densities. It is also characterized by the presence of both Dirac and Haldane gaps.
While our qualitative conclusion seems rather rigorous, this study is insufficient to establish the
precise structure of the phase diagram in the plane of the applied electric and magnetic fields. It would
be very interesting to clarify the details of the corresponding phase diagram either experimentally or
numerically.

It may be instructive to note that the thermal broadening of Landau levels plays a relatively
important technical role in our analysis and in the description of the electrified homogeneous
phases. Indeed, by using a nonzero temperature, we were able to unambiguously describe the
surface ground states with adjustable partial fillings of Landau levels, needed to screen the
external electric field. Certainly, the corresponding ground states allow a well defined zero
temperature limit, but their description may become more subtle. By noting that surface
impurities also broaden Landau levels, we suggest that their presence could lead to a
realization of the electrified magnetic catalysis similar to that in
Sec.~\ref{sec:gap-equation-Coulomb-results-kappa}.

As is clear from our study, the low-energy model for the surface states of 3D TIs is essentially
a (2+1)-dimensional QED, supplemented by certain constraints. 
The generation of different types of gaps, such as those describing spontaneous parity breaking 
and chiral symmetry breaking were studied in QED$_{2+1}$ without background
electromagnetic fields in Refs.~\cite{Appelquist:1986,Appelquist1:1986,Wijewardhana} a long time
ago. Moreover, it was shown that the Dirac mass can be spontaneously generated, while the
Haldane mass is energetically disfavored \cite{Appelquist1:1986}. Clearly, this is not the
case in the problem at hand, where both types of gaps are generated on the TI surfaces. This
is due to the fact that the TR and inversion symmetries are explicitly broken by the external
magnetic and electric fields. Furthermore, we find that the Haldane gap dominates at small values
of the electric field. This situation is reminiscent of the dynamically enhanced Zeeman splitting
in graphene \cite{Gorbar:2011kc}.

\section{Conclusion}
\label{sec:Conclusion}

In this study, we considered the dynamical generation of gaps in a slab of a 3D TI, such as Bi$_2$Se$_3$,
placed in the magnetic and electric fields perpendicular to its surfaces. (Although we used the model
parameters for Bi$_2$Se$_3$, the main conclusions should be valid for all similar TIs.) Note
that the conducting states on the TI slab surfaces and the overall geometry of the system are rather
similar to bilayer graphene. On the other hand, the degeneracy connected with the valley and spin
degrees of freedom, which is responsible for a variety of quantum Hall states in bilayer graphene,
is absent in a TI slab. Still, there are notable similarities in the dynamics of gap generation in
these two physical systems. For example, the valley quantum Hall (or layer polarized) state, which is
realized in a sufficiently strong external electric field in bilayer graphene, resembles the homogeneous
state considered in this paper.

By solving the gap equations for the surface quasiparticle propagators in a simple model with
short- and long-range interactions, we found that {\em both} the Dirac and Haldane gaps are dynamically
generated in the TI slab in external electric and magnetic fields. The underlying mechanism is a different
version of the magnetic catalysis. Because of a large surface dielectric constant, the Coulomb
interaction appears to play a minor role in the dynamics. Unlike the Dirac gap, the Haldane gap
respects the $U(2)$ symmetry with the generators given in Eq.~(\ref{gap-Coulomb-results-generators}),
but breaks the parity and TR symmetries. Since both discrete symmetries are
explicitly broken by external electric and magnetic fields, the generation of the Haldane gap does
not break any symmetries. The Dirac gap, on the other hand, is generated only in the
presence of an electric field. The result of such an {\em electrified} magnetic catalysis is a spontaneous
breaking of the $U(2)$ symmetry.

By comparing our results with the findings in Ref.~\cite{Vishwanath:2010}, we argued that the
homogeneous phase with dynamically generated Dirac and Haldane gaps is the true ground state
in the TI slab in nonzero magnetic and sufficiently strong electric fields. The precise structure of
the phase diagram in the plane of applied electric and magnetic fields remains to be clarified,
however.

\acknowledgments

The authors are grateful to V.P. Gusynin for useful discussions.
The work of E.V.G. was supported partially by the Ukrainian State Foundation for Fundamental
Research. The work of V.A.M. and P.O.S. was supported by the Natural Sciences and Engineering
Research Council of Canada. The work of I.A.S. was supported in part by the
U.S. National Science Foundation under Grant No.~PHY-1404232.

\appendix

\section{Wave functions and fermion propagator}
\label{sec:wf-Green}

In this Appendix, we determine the wave functions and the fermion propagator in the model with the free
Hamiltonian in Eq.~(\ref{model-H-s-0}). For the simplicity of notation, we will drop the superscript $(0)$ in
$\mu_{\lambda}^{(0)}$ and $m_{\lambda}^{(0)}$  in this Appendix.

\subsection{Wave functions}
\label{sec:wave-function}

The eigenvalue problem $H^{(0)}_{\lambda }\psi=\omega\psi$ for the model Hamiltonian in Eq.~(\ref{model-H-s-0})
reduces to the following equation:
\begin{equation}
\left(
                        \begin{array}{cc}
                          m_{\lambda }-\mu_{\lambda} & v_F(\partial_x+s_Bk_y+eBx) \\
                          v_F(-\partial_x+s_Bk_y+eBx) & -m_{\lambda}-\mu_{\lambda } \\
                        \end{array}
                      \right)\phi(x) =\omega\phi(x),
\label{wf-system-0}
\end{equation}
after we choose the wave function in the form $\psi(\mathbf{r})=e^{is_Bk_y y}\phi(x)$, where
$\phi(x)=\left( \phi_1(x),\phi_2(x)\right )^{T}$ and $s_{B}=\sign{(eB)}$. It terms of the new variable,
\begin{equation}
\xi =\sqrt{|eB|}\left(\frac{k_y}{|eB|}+x\right),
\label{wf-notations}
\end{equation}
Eq.~(\ref{wf-system-0}) can be rewritten in the form
\begin{eqnarray}
\label{wf-system-1-be}
&&\left(m_{\lambda}-\mu_{\lambda }-\omega\right)\phi_1(\xi) +v_F \sqrt{|eB|} \left(\partial_{\xi} +s_B\xi\right)\phi_2(\xi)=0, \\
\label{wf-system-1-ee}
&&v_F \sqrt{|eB|} \left(-\partial_{\xi} +s_B\xi\right)\phi_1(\xi)-\left(m_{\lambda}+\mu_{\lambda }+\omega\right)\phi_2(\xi)=0.
\end{eqnarray}
For simplicity, here we consider only the case with $s_{B}=+1$. The solutions to this system of equations take the form:
\begin{eqnarray}
\phi_1(\xi) &=& \frac{v_F\sqrt{2|eB|}}{\mu_{\lambda }+\omega-m_{\lambda}} \frac{p^2}{2} D_{p^2/2-1}
\left(\sqrt{2}\xi\right), \\
\quad \phi_2(\xi) &=& D_{p^2/2} \left(\sqrt{2}\xi\right),
\label{wf-sbp-phi12}
\end{eqnarray}
where $p= \sqrt{(\omega+\mu_{\lambda })^2-m_{\lambda }^2}/(v_F\sqrt{|eB|})$ and $D_{p^2/2}(\sqrt{2} \xi)$
is the parabolic cylinder function \cite{Bateman}. By requiring that solutions are finite at $|\xi|\rightarrow\infty$,
we find the quantization condition $p^2/2=n$, where $n$ is a nonnegative integer. In this special case, the
parabolic cylinder functions can be expressed in terms of Hermitian polynomials $H_{n}(\xi)$, i.e.,
\begin{equation}
D_{n}(\sqrt{2}\xi) = \frac{e^{-\xi^2/2}H_{n}(\xi)}{\sqrt{2^n}}.
\label{wf-parabolic-hermit}
\end{equation}
Thus, the final expressions for the normalized wave functions take the form:
\begin{eqnarray}
\label{wf-sbp}
\quad \psi_{s_B=+1, n}(\mathbf{r}) =  \frac{1}{\sqrt{2 l}}
\sqrt{\frac{\omega_n+\mu_{\lambda }-m_{\lambda}}{\omega_n+\mu_{\lambda }}} e^{ik_y y}\left(
                                                 \begin{array}{c}
                                                   \frac{v_F\sqrt{2n|eB|}}{\mu_{\lambda}+\omega_n-m_{\lambda}} Y_{n-1}(\xi) \\
                                                   Y_{n}(\xi)  \\
                                                 \end{array}
                                               \right),
\end{eqnarray}
where $Y_n(\xi)=\frac{e^{-\xi^2/2} H_n(\xi)}{\sqrt{2^n n! \sqrt{\pi}}}$ and $l=1/\sqrt{|eB|}$ is the 
magnetic length. Note that the normalized wave function in the case of $s_B=-1$ is given by 
$\psi_{s_B=-1, n}(\mathbf{r})=\left. (-i\sigma_y)\psi^{*}_{s_B=+1, n}(\mathbf{r})\right|_{m_{\lambda}\to-m_{\lambda}}$.
The corresponding energy eigenvalues are
\begin{equation}
\omega_{n=0}=-\mu_{\lambda }-s_Bm_{\lambda}, \qquad
\omega_{n>0}=-\mu_{\lambda }\pm M_n,
\label{wf-spectrum}
\end{equation}
where $M_n=\sqrt{m^2_{\lambda }+n\epsilon_B^2}$ and $\epsilon_{B}= \sqrt{2v_F^2|eB|}$ is the Landau energy scale.

\subsection{Fermion propagator}
\label{sec:Greens-function}

In this subsection we derive the free fermion propagator in the model under consideration. In the mixed
frequency-coordinate space representation, the fermion (Feynman) propagator is formally defined by
\begin{equation}
S(\omega, \mathbf{r}, \mathbf{r}^{\prime}) \equiv i\sum_{n}
\frac{\psi_n(\mathbf{r})\bar{\psi}_n(\mathbf{r})}{\omega-\omega_n+i0\sign{(\omega)}},
\label{Green-def}
\end{equation}
where the sum runs over the complete set of quasiparticle eigenstates given by Eq.~(\ref{wf-spectrum}).

It is convenient to start with the derivation of the lowest Landau level (LLL) contribution to the fermion
propagator, i.e.,
\begin{equation}
S_{n=0}(\omega, \mathbf{r}, \mathbf{r}^{\prime}) =  i \int \frac{dk_y}{2\pi} \frac{\psi_{n=0}(\mathbf{r})
\bar{\psi}_{n=0}(\mathbf{r}^{\prime})}{\omega+\mu_{\lambda }+s_Bm_{\lambda}+i0\sign{(\omega)}}
= -is_B \frac{P_{-}}{2\pi l^2} \frac{e^{i\Phi(\mathbf{r}, \mathbf{r}^{\prime})-\eta/2}}{\omega+\mu_{\lambda }+s_Bm_{\lambda}
+i0\sign{(\omega)}},
\label{Green-n=0-app}
\end{equation}
where $P_{\pm} =\left(1 \pm s_B \gamma^0\right)/2$,
$\Phi(\mathbf{r}, \mathbf{r}^{\prime})= -eB(x+x^{\prime})(y-y^{\prime})/2$ is the Schwinger phase,
and $\eta= (\mathbf{r}-\mathbf{r}^{\prime})^2/(2l^2)$.

The contribution of higher Landau levels to the fermion propagator is given by
\begin{eqnarray}
S_{n>0}(\omega, \mathbf{r}, \mathbf{r}^{\prime})&=& \sum_{n=1}^{\infty}\frac{ie^{i\Phi(\mathbf{r}, \mathbf{r}^{\prime})-\eta/2}}{2\pi l^2} \Bigg\{\frac{s_B\left(\omega+\mu_{\lambda }\right)\left\{ L_{n-1}(\eta)P_{+} -L_{n}(\eta)P_{-}\right\}}
{\left(\omega+\mu_{\lambda }+i0\sign{(\omega)}\right)^2-M_n^2}\nonumber\\
&+&\frac{m_{\lambda}\left(L_{n-1}(\eta)P_{+} +L_{n}(\eta)P_{-}\right) -i\frac{v_F}{l^2}L_{n-1}^1(\eta)
\left(\bm{\gamma}\cdot(\mathbf{r}-\mathbf{r}^{\prime})\right) }{\left(\omega+\mu_{\lambda }+i0\sign{(\omega)}\right)^2-M_n^2} \Bigg\},
\label{Green-ng0-app}
\end{eqnarray}
where $L^j_{n} \left(x\right)$ are the generalized Laguerre polynomials (by definition $L_n\equiv L_n^0$). In the derivation, we performed
the summation over the quasiparticle energies and integrated over $k_y$ by using formula
7.377 in Ref.~\cite{Gradshtein}.

In the case of a nonzero temperature, the energies are replaced by the Matsubara frequencies, i.e.,
$\omega\to i\omega_{m^{\prime}}=i\pi T(2m^{\prime}+1)$. In the gap equation (\ref{gapEq-Coulomb-gap-1}),
the corresponding propagators enter in the form of a sum over Matsubara frequencies. The corresponding
results for the sums are
\begin{eqnarray}
\label{Green-T-n=0-app}
T\sum_{m^{\prime}=-\infty}^{\infty} S_{n=0}(i\omega_{m^{\prime}}, \mathbf{r}, \mathbf{r}^{\prime})
&=&\frac{s_BP_{-} e^{i\Phi(\mathbf{r}, \mathbf{r}^{\prime})-\eta/2}}{2\pi l^2} \frac{1-2n_{F}\left(\mu_{\lambda }+s_B m_{\lambda }\right)}{2},\\
\label{Green-T-ng0-app}
T\sum_{m^{\prime}=-\infty}^{\infty}S_{n>0}(i\omega_{m^{\prime}}, \mathbf{r}, \mathbf{r}^{\prime})&=& -\frac{e^{i\Phi(\mathbf{r}, \mathbf{r}^{\prime})-\eta/2}}{4\pi l^2} \sum_{n=1}^{\infty} \Bigg\{ \left[n_{F}\left(M_n-\mu_{\lambda }\right)
-n_{F}\left(M_n+\mu_{\lambda }\right)\right] \left[ L_{n-1}(\eta)P_{+} -L_{n}(\eta)P_{-}\right] \nonumber\\
&+&\left[m_{\lambda}\left(L_{n-1}(\eta)P_{+} +L_{n}(\eta)P_{-}\right) -i\frac{v_F}{l^2}L_{n-1}^1(\eta)
\left(\bm{\gamma}\cdot(\mathbf{r}-\mathbf{r}^{\prime})\right) \right] \nonumber\\
&\times&\frac{n_{F}\left(M_n+\mu_{\lambda }\right)+n_{F}\left(M_n-\mu_{\lambda }\right)-1}{M_n} \Bigg\}.
\end{eqnarray}
where $n_{F}(x)=1/\left(e^{x/T}+1\right)$ is the Fermi-Dirac distribution function.

\subsection{Two stripes separated by a domain wall}
\label{sec:domain-wall}

In this subsection, we consider the Dirac problem with the inhomogeneous gap in the form $m(x)=|m|\sign{(x)}$,
which models an infinitely thin transition region at $x=0$ that separates two wide stripes of phases with
masses of opposite signs. For the sake of simplicity, we set $\lambda=+1$ and $s_{B}=+1$ in this subsection.

By making use of wave functions found in Appendix~\ref{sec:wave-function}, we obtain the following
solutions in the regions with positive and negative gaps:
\begin{eqnarray}
\label{stripe-zero-modes-wf-sbp-phi12-mg0}
&&x>0:
\quad
\phi_1(\xi) = \frac{v_F\sqrt{2|eB|}}{\mu+\omega-|m|} \frac{p^2}{2} D_{\frac{p^2}{2}-1}\left(\sqrt{2}\xi\right),
\quad
\phi_2(\xi)=D_{\frac{p^2}{2}}\left(\sqrt{2}\xi\right),\\
\label{stripe-zero-modes-wf-sbp-phi12-ml0}
&&x<0:
\quad
\phi_1(\xi) = -\frac{v_F\sqrt{2|eB|}}{\mu+\omega+|m|} \frac{p^2}{2} D_{\frac{p^2}{2}-1}\left(-\sqrt{2}\xi\right),
\quad
\phi_2(\xi)=D_{\frac{p^2}{2}}\left(-\sqrt{2}\xi\right),
\end{eqnarray}
where we assumed that the corresponding wave functions must vanish at $x\to\pm\infty$. By matching the wave
functions at $x=0$, i.e.,
\begin{equation}
C_{+}\psi_{x>0}\Big|_{x=0}=C_{-}\psi_{x<0}\Big|_{x=0},
\label{stripe-zero-modes-matching}
\end{equation}
we derive the following spectral equation:
\begin{equation}
\frac{D_{\frac{p^2}{2}-1}\left(\sqrt{2}k_yl\right)}{D_{\frac{p^2}{2}}\left(\sqrt{2}k_yl\right)}
\frac{1}{\omega+\mu-|m|} = -
\frac{D_{\frac{p^2}{2}-1}\left(-\sqrt{2}k_yl\right)}{D_{\frac{p^2}{2}}\left(-\sqrt{2}k_yl\right)}
\frac{1}{\omega+\mu+|m|}.
\label{stripe-zero-modes-spectral-eq}
\end{equation}
Note that this spectral equation looks somewhat similar to Eq.~(2.19) for bound states in Ref.~\cite{Shen:book}. We note,
however, that the wave function in our case decreases at $|x|\to\infty$ polynomially rather than exponentially.

By making use of the wave functions, we can also give the following explicit results for the density of charge
carriers (plus sign) and the chiral condensate (minus sign):
\begin{eqnarray}
\mbox{tr}[(\gamma^0)^{\frac{1\pm 1}{2}}G(u,u)]&=& \int \frac{dk_y}{2\pi} \sum_{n} \frac{\sign{\left[\omega_n(k_y)\right]}}{2}
|C_{+}\left[\omega_n(k_y)+\mu, k_{y}\right]|^2\Bigg\{ \theta{(x)}\left[\frac{(\omega_n(k_y)+\mu+|m|)^2}{\epsilon_B^2} \left|D_{\nu_{n}-1}
\left(\sqrt{2}k_{y}l+\sqrt{2}\frac{x}{l}\right)\right|^2 \right. \nonumber\\
&\pm&\left.\left|D_{\nu_{n}}\left(\sqrt{2}k_{y}l+\sqrt{2}\frac{x}{l}\right)\right|^2\right] +\theta{(-x)}\left|\frac{D_{\nu_{n}}
\left(\sqrt{2}k_yl\right)}{D_{\nu_{n}}\left(-\sqrt{2}k_yl\right)}\right|^2 \nonumber\\
&\times&\left[\frac{(\omega_n(k_y)+\mu-|m|)^2}{\epsilon_B^2} \left|D_{\nu_{n}-1}\left(-\sqrt{2}k_{y}l-\sqrt{2}\frac{x}{l}\right)\right|^2
\pm \left|D_{\nu_{n}}\left(-\sqrt{2}k_{y}l-\sqrt{2}\frac{x}{l}\right)\right|^2\right]\Bigg\},
\label{stripe-zero-modes-gap-m0}
\end{eqnarray}
where $\omega_n(k_y)$ denotes the roots of Eq.~(\ref{stripe-zero-modes-spectral-eq}),
$\nu_{n}=\left[\left(\omega_n(k_y)+\mu\right)^2-m^2\right]/\epsilon_B^2$, and $C_{+}(\omega, k_{y})$ is a
normalization constant. Note that contrary to the homogeneous phase, $\omega_n(k_y)$ now explicitly depends on $k_y$.

\section{Effective action}
\label{sec:App-action-true}

In this appendix we derive the one-loop Baym--Kadanoff (BK) effective action for the full
surface quasiparticle propagators $G_{\lambda}$. We begin with the part of the effective
action connected with the Coulomb interaction. (For a similar derivation in bilayer graphene, see
Ref.~\cite{bilayer}.) The interaction Hamiltonian, $H_{\rm int}$, of the
Coulomb interaction in 3D has the standard form
\begin{equation}
H_{\rm int} = \frac{e^2}{8\pi \epsilon_0 \kappa}\int d^2\mathbf{r}dz\,d^2\mathbf{r}^{\prime}dz^{\prime}\,
\frac{n(\mathbf{r},z)n(\mathbf{r}^{\prime},z^{\prime})}{\sqrt{(\mathbf{r}-\mathbf{r}^{\prime})^2+(z-z^{\prime})^2}},
\label{app-action-H-int-z}
\end{equation}
where $\epsilon_0\approx8.854\times10^{-12}~\mbox{F/m}$ is the vacuum dielectric constant and $\kappa$ is a dielectric permittivity.
Since we wish to consider a TI slab in an external electric field, the density of charge carriers in the system under consideration consists of
four terms:
\begin{equation}
n(\mathbf{r},z) = \delta(z-L_z)n_{c, -}+\delta(z-l_z)n_{-}(\mathbf{r})+\delta(z+l_z)n_{+}(\mathbf{r})+\delta(z+L_z)n_{c,+},
\label{app-action-n}
\end{equation}
where $n_{c, \lambda}$ denote the densities of charge carriers on the capacitor plates separated by the distance $2L_z$ (clearly, we assume that
these charge densities are uniform and do not depend on $\mathbf{r}$ in order to produce constant electric field in which our TI
slab is situated), and $n_{\lambda}(\mathbf{r})=\psi^{\dagger}_{\lambda}(\mathbf{r})\psi_{\lambda}(\mathbf{r})$ denotes the density of charge
carriers on the surfaces of the slab whose width is $2l_z$. Obviously, we assume that $L_z > l_z$.

By using Eq.~(\ref{app-action-n}), we can easily integrate over $z$ in Eq.~(\ref{app-action-H-int-z}). The result reads as
\begin{eqnarray}
H_{\rm int} &=& \frac{1}{2}\int d^2\mathbf{r}d^2\mathbf{r}^{\prime}\,
\Bigg\{U(\mathbf{r}-\mathbf{r}^{\prime})\left[n_{+}(\mathbf{r})n_{+}(\mathbf{r}^{\prime}) +n_{-}(\mathbf{r})n_{-}(\mathbf{r}^{\prime})\right]
+2U_{\rm s, inter}(\mathbf{r}-\mathbf{r}^{\prime})n_{+}(\mathbf{r})n_{-}(\mathbf{r}^{\prime})
\nonumber\\
&+&2U_{\rm sc, 1}(\mathbf{r}-\mathbf{r}^{\prime})\left[n_{+}(\mathbf{r})n_{c,+} +n_{-}(\mathbf{r})n_{c,-}
\right]
+2U_{\rm sc, 2}(\mathbf{r}-\mathbf{r}^{\prime})\left[n_{+}(\mathbf{r})n_{c,-} +n_{-}(\mathbf{r})n_{c,+}
\right]\nonumber\\
&+&\kappa_{\rm surf}U(\mathbf{r}-\mathbf{r}^{\prime})\left[n_{c,+}n_{c,+}+n_{c,-}n_{c,-}\right]+2U_{\rm c, inter}(\mathbf{r}-\mathbf{r}^{\prime})n_{c,-}n_{c,+} \Bigg\},
\label{app-action-H-int-n}
\end{eqnarray}
where the corresponding Coulomb potentials are
\begin{eqnarray}
\label{app-action-U-be}
U(\mathbf{r})&=&\frac{e^2}{4\pi \epsilon_0 \kappa_{\rm surf}} \frac{1}{r}, \quad
U_{\rm s, inter}(\mathbf{r}) = \frac{e^2}{4\pi \epsilon_0 \kappa_{\rm surf}} \frac{1}{\sqrt{r^2+4l_z^2}}, \quad
U_{\rm sc, 1}(\mathbf{r}) =\frac{e^2}{4\pi \epsilon_0 \kappa_{\rm surf}} \frac{1}{\sqrt{r^2+|L_z-l_z|^2}}, \nonumber\\
U_{\rm sc, 2}(\mathbf{r}) &=& \frac{e^2}{4\pi \epsilon_0 \kappa_{\rm surf}} \frac{1}{\sqrt{r^2+|L_z+l_z|^2}}, \quad
U_{\rm c, inter}(\mathbf{r}) = \frac{e^2}{4\pi \epsilon_0} \frac{1}{\sqrt{r^2+4L_z^2}}.
\label{app-action-U-ee}
\end{eqnarray}
The physical meaning of Eq.~(\ref{app-action-H-int-n}) is transparent. Its first term describes the standard Coulomb interaction of
quasiparticles on the top and bottom surfaces of the TI slab. The second term corresponds to the intersurface interaction, therefore,
$U_{\rm s, inter}$ contains additional term $4l_z^2$ in the denominator compared to $U$. The second line in Eq.~(\ref{app-action-H-int-n})
describes interactions of surface quasiparticles with the charge densities on the capacitor plates. The last line in
Eq.~(\ref{app-action-H-int-n}) does not depend on $n_{\lambda}(\mathbf{r})$ and is, therefore, irrelevant for the gap equations (obviously,
this line describes the electrostatic interaction of charge densities on capacitor plates). The Fourier transforms of interactions
(\ref{app-action-U-ee}) are given by
\begin{eqnarray}
\label{app-action-U-F-be}
U(\mathbf{k}) &=& \frac{e^2}{2 \epsilon_0 \kappa_{\rm surf}} \frac{1}{k}, \quad
U_{\rm s, inter}(\mathbf{k}) = \frac{e^2}{2 \epsilon_0 \kappa_{\rm surf}}  \frac{e^{-2l_zk}}{k}, \quad
U_{\rm sc, 1}(\mathbf{k}) = \frac{e^2}{2 \epsilon_0 \kappa_{\rm surf}}  \frac{e^{-|L_z-l_z|k}}{k}, \nonumber\\
U_{\rm sc, 2}(\mathbf{k}) &=& \frac{e^2}{2 \epsilon_0 \kappa_{\rm surf}}  \frac{e^{-|L_z+l_z|k}}{k}, \quad
U_{\rm c, inter}(\mathbf{k}) = \frac{e^2}{2 \epsilon_0}  \frac{e^{-2L_zk}}{k},
\label{app-action-U-F-ee}
\end{eqnarray}
where $k=|\mathbf{k}|$. Further, it is convenient to rewrite the first term in Eq.~(\ref{app-action-H-int-n}) as follows:
\begin{equation}
U(\mathbf{r}-\mathbf{r}^{\prime})\left[n_{+}(\mathbf{r})n_{+}(\mathbf{r}^{\prime}) +n_{-}(\mathbf{r})n_{-}(\mathbf{r}^{\prime})\right]
=U(\mathbf{r}-\mathbf{r}^{\prime})\Psi^{\dag}(\mathbf{r})\Psi(\mathbf{r})\Psi^{\dag}(\mathbf{r}^{\prime})\Psi(\mathbf{r}^{\prime})
-2U(\mathbf{r}-\mathbf{r}^{\prime})n_{+}(\mathbf{r})n_{-}(\mathbf{r}^{\prime}).
\end{equation}

The one-loop BK effective action \cite{BK} in the model under consideration reads as
\begin{eqnarray}
\Gamma(G) &=& -i\,\sum_{\lambda=\pm}\mathrm{Tr}\left[\mathrm{Ln}G^{-1}_{\lambda}+S^{-1}_{\lambda}G_{\lambda} -1\right]
+\frac{e^2}{2}\int d^3u\int d^3u^{\prime} \Bigg\{\sum_{\lambda=\pm}\mathrm{tr}\left[\gamma^0G_{\lambda}(u,u^{\prime})
\gamma^0G_{\lambda}(u^{\prime},u)\right]D(u^{\prime}-u)
\nonumber\\
&-& \sum_{\lambda=\pm}\mathrm{tr}\left[\gamma^0G_{\lambda}(u,u)\right]\sum_{\lambda^{\prime}=\pm}\mathrm{tr}
\left[\gamma^0G_{\lambda^{\prime}}(u^{\prime},u^{\prime})\right]D(u^{\prime}-u) \nonumber\\
&+&2\sum_{\lambda=\pm}\mathrm{tr}
\left[\mathcal{P}_{+}^{\lambda}\gamma^0G_{\lambda}(u,u^{\prime})\mathcal{P}_{-}^{\lambda}\gamma^0G_{\lambda}(u^{\prime},u)\right] \left[D_{\rm s,inter}(u^{\prime}-u)-D(u^{\prime}-u)\right]
\nonumber\\
&-&2\sum_{\lambda=\pm}\mathrm{tr}\left[\mathcal{P}_{+}^{\lambda}\gamma^0G_{\lambda}(u,u)\right]\sum_{\lambda^{\prime}=\pm}\mathrm{tr}
\left[\mathcal{P}_{-}^{\lambda^{\prime}}\gamma^0G_{\lambda^{\prime}}(u^{\prime},u^{\prime})\right]
\left[D_{\rm s,inter}(u^{\prime}-u)-D(u^{\prime}-u)\right] \nonumber\\
&+&2\sum_{\lambda=\pm}\mathrm{tr}\left[\mathcal{P}_{+}^{\lambda}\gamma^0G_{\lambda}(u,u)n_{c,+}+\mathcal{P}_{-}^{\lambda}\gamma^0G_{\lambda}(u,u)n_{c,-}\right]D_{\rm sc,1}(u^{\prime}-u)
\nonumber\\ &+&2\sum_{\lambda=\pm}\mathrm{tr}\left[\mathcal{P}_{+}^{\lambda}\gamma^0G_{\lambda}(u,u)n_{c,-}+\mathcal{P}_{-}^{\lambda}\gamma^0G_{\lambda}(u,u)n_{c,+}\right]
D_{\rm sc,2}(u^{\prime}-u)-\left[n_{c,+}n_{c,+}+n_{c,-}n_{c,-}\right]\kappa_{\rm surf}D(u^{\prime}-u)
\nonumber\\
&-&2n_{c,+}n_{c,-}D_{\rm c,inter}(u^{\prime}-u)\Bigg\} -\frac{G_{\rm int}}{2}\int d^3u \sum_{\lambda=\pm}\Big( \tr{\big[\gamma^0G_{\lambda}(u,u)\big]}\tr{\big[\gamma^0G_{\lambda}(u,u)\big]}
-\tr{\big[\gamma^0G_{\lambda}(u,u)\gamma^0G_{\lambda}(u,u)\big]}
\Big), \nonumber\\
\label{app-action-BK}
\end{eqnarray}
where $\mathcal{P}_{\pm}^{\lambda}=(1\pm\lambda)/2$ are the surface projectors. The trace in the first term is taken in the functional sense, the
trace in the rest of terms is taken over spinor indices.

The extremum of the effective action $\frac{\delta \Gamma(G)}{\delta G_{\lambda}} =0$ defines the following Schwinger-Dyson equation for the
full fermion propagator:
\begin{eqnarray}
iG^{-1}_{\lambda}(u,u^\prime) &=& iS^{-1}_{\lambda}(u,u^\prime) - e^2 \Bigg\{ \gamma^0 G_{\lambda}(u,u^{\prime}) \gamma^0 D(u^{\prime}-u)
-\gamma^0 \sum_{\lambda=\pm}\mathrm{tr}\left[\gamma^0G_{\lambda}(u,u)\right]\delta^3(u-u^{\prime})\tilde{D}(0) \nonumber\\
&-& \gamma^0 \mathrm{tr}\left[\gamma^0 G_{-\lambda}(u,u)\right] \delta^3(u-u^{\prime})\left[\tilde{D}_{\rm s,inter}(0)-\tilde{D}(0)\right]
\nonumber\\
&+&\gamma^0 n_{c,\lambda})\delta^3(u-u^{\prime})\tilde{D}_{\rm sc,1}(0)
+\gamma^0 n_{c,-\lambda}\delta^3(u-u^{\prime})\tilde{D}_{\rm sc,2}(0)\Bigg\} \nonumber\\
&-&G_{\rm int}
\left\{ \gamma^0 G_{\lambda}(u,u) \gamma^0 - \gamma^0\, \mbox{tr}[\gamma^0G_{\lambda}(u,u)]\right\}\delta^{3}(u-u^{\prime}),
\label{app-action-gap}
\end{eqnarray}
where the contribution due to the third term in the curly brackets in Eq.~(\ref{app-action-BK})
is zero because the fermion propagator of the full model is diagonal in surface indices. The trace is taken
over the spinor indices only, $\tilde{D}(0)$, $\tilde{D}_{\rm s, inter}(0)$, $\tilde{D}_{\rm sc,1}(0)$,
$\tilde{D}_{\rm sc,2}(0)$ are the Fourier transforms of the corresponding interactions at zero momentum, and
\begin{eqnarray}
\label{app-action-D-F-be}
D(u) &\approx& \delta(t)
\frac{1}{4\pi\epsilon_0\kappa_{\rm surf}} \int \frac{dk}{2\pi} J_0(kr),
\label{app-action-D-F-ee}
\end{eqnarray}
where $J_0(kr)$ is the Bessel function. The overall neutrality condition (\ref{gapEq-Coulomb-Hartree}) implies that the second term in the curly brackets in
Eq.~(\ref{app-action-gap}) is equal to zero. The last two terms in the curly brackets for $n_{c,+}=-n_{c,-}$
(which stems from the symmetric charge distribution on the opposite surfaces of the
slab) are equal to
\begin{equation}
\left[n_{c,\lambda}\tilde{D}_{\rm sc,1}(0)+ n_{c,-\lambda}\tilde{D}_{\rm sc,2}(0)\right]=-\gamma^0 \frac{1}{2\epsilon_0\kappa_{\mathrm{surf}}}
n_{c,\lambda}\left[(L_z-l_z)- (L_z+l_z)\right]=n_{c,\lambda}\gamma^0 \frac{l_z}{\epsilon_0\kappa_{\mathrm{surf}}}.
\label{app-action-capacitor-term}
\end{equation}
Notice the fact that the dependence on $L_z$ cancels out in Eq.~(\ref{app-action-capacitor-term}).
This means that the formalism correctly describes the TI slab in an applied external electric
field. Thus, the gap equation (\ref{app-action-gap}) takes the form
\begin{eqnarray}
iG^{-1}_{\lambda}(u,u^\prime) &=& iS^{-1}_{\lambda}(u,u^\prime) - e^2 \left\{ \gamma^0 G_{\lambda}(u,u^{\prime}) \gamma^0 D(u^{\prime}-u)
-\gamma^0\frac{l_z\delta^3(u-u^{\prime})}{\epsilon_0\kappa_{\rm surf}} \left[n_{-\lambda}
+ n_{c, -\lambda}\right]\right\} \nonumber\\
&-&G_{\rm int}
\left\{ \gamma^0 G_{\lambda}(u,u) \gamma^0 - \gamma^0\, \mbox{tr}[\gamma^0G_{\lambda}(u,u)]\right\}\delta^{3}(u-u^{\prime}).
\label{app-action-gap-simple}
\end{eqnarray}
The last term in the first curly brackets has a clear physical meaning. It describes a superposition of electric fields due to the
capacitor plates and charged surfaces of the TI. Taking into account condition (\ref{model-DOS-zero-field}), i.e.,
$n_{\lambda}=-\lambda \epsilon_0\mathcal{E}/e$, we conclude that the corresponding term in the gap equation (\ref{app-action-gap}) vanishes
and the gap equation takes the following final form:
\begin{equation}
iG^{-1}_{\lambda}(u,u^\prime) = iS^{-1}_{\lambda}(u,u^\prime) - e^2  \gamma^0 G_{\lambda}(u,u^{\prime}) \gamma^0 D(u^{\prime}-u)  -G_{\rm int}
\left\{ \gamma^0 G_{\lambda}(u,u) \gamma^0 - \gamma^0\, \mbox{tr}[\gamma^0G_{\lambda}(u,u)]\right\}\delta^{3}(u-u^{\prime}). \nonumber\\
\label{app-action-gap-fin}
\end{equation}
It is worth noting that we are working in the grand canonical ensemble. The corresponding free energy
density is expressed through the effective action $\Gamma$ as $\Omega=-\Gamma/(VT)$, where
$VT$ is a $(2+1)$ space-time volume. On the solution of the gap equation (\ref{app-action-gap-fin}) we
have the following surface free energy density (compare with Appendix C in Ref.~\cite{Gorbar:2008hu}):
\begin{eqnarray}
\label{app-free-energy-T}
\Omega_{\lambda} &=&- \frac{1}{8\pi l^2} \left[1-2n_{F}\left(\Delta_{\rm eff, \lambda}\right)\right]
 \left[\Delta_{\rm eff, \lambda}+\mu^{(0)}_{\lambda }+s_Bm^{(0)}\right]  -\frac{1}{4\pi l^2}\sum_{n=1}^{\infty} \Bigg\{ (\mu^{(0)}_{\lambda }+\mu_{n,\lambda})
\left[n_{F}\left(M_n-\mu_{n,\lambda }\right)-n_{F}\left(M_n+\mu_{n,\lambda }\right)\right]
\nonumber\\
&+&\frac{2M_n^2+m_{n,\lambda }(m^{(0)}-m_{n,\lambda })}{M_n}
\left[1-n_{F}\left(M_n+\mu_{n,\lambda }\right)-n_{F}\left(M_n-\mu_{n,\lambda }\right)\right] \Bigg\} -\frac{\epsilon_0\mathcal{E}^2}{2\kappa_{\rm surf}}\left(\kappa_{\rm surf}L_z-l_z\right),
\end{eqnarray}
where the last term has a transparent physical meaning and represents the energy density of an electric field outside the slab.

\section{The kernel coefficients $\mathcal{K}^{(0)}_{m,n}$}
\label{sec:app-K}

In this appendix, we give the definition for the kernel coefficients $\mathcal{K}^{(0)}_{m,n}$ used in Coulomb part of the gap equation in the static approximation. Further, we provide an explicit expression for those coefficients in the case without polarization effects.
The kernel coefficients are defined by \cite{Gorbar:2011kc}
\begin{eqnarray}
\label{app-K-def}
\mathcal{K}^{(0)}_{m,n} &=&\int_0^{\infty} \frac{dk}{2\pi} \frac{kl\mathcal{L}^{(0)}_{m,n}(kl)}{k+\Pi(0,k)}, \\
\label{app-K-L-0}
\mathcal{L}^{(0)}_{m,n}&=&\frac{1}{l^2} \int_0^{\infty} dr \,r\,e^{-\frac{r^2}{2l^2}}L_m^0\left(\frac{r^2}{2l^2}\right)L_n^0
\left(\frac{r^2}{2l^2}\right)J_0(kr) =(-1)^{m+n}e^{-\frac{k^2l^2}{2}}L_m^{n-m}\left(\frac{k^2l^2}{2}\right)L_n^{m-n}\left(\frac{k^2l^2}{2}
\right),
\label{L0mn}
\end{eqnarray}
where we used formula 7.422.2 in Ref.~\cite{Gradshtein} in order to perform the integration in Eq.~(\ref{L0mn}).

By neglecting the polarization effects, i.e., setting $\Pi(0,k)=0$, and using formula 2.19.14.15 in Ref.~\cite{Prudnikov}, one
obtains the following explicit result:
\begin{equation}
\mathcal{K}^{(0)}_{m,n}\Big|_{\Pi\to0} = \int_0^{\infty}
\frac{dx}{2\pi} \int_0^{\infty} dt\, e^{-t} L_{m}^{0}(t)L_{n}^{0}(t) J_0(x\sqrt{2t})
=\frac{\Gamma(n+1/2)\Gamma(m+1/2)}{2\sqrt{2}\,\pi^{3/2}\,m!\,n!} {}_3F_{2}(-m,-n,1/2;1/2-m,1/2-n;1) ,
\label{app-K-Pi0}
\end{equation}
where ${}_3F_{2}$ is the hypergeometric function.

\end{document}